\documentclass{mn2e}
\usepackage{graphics}
\usepackage{psfrag}
\usepackage{xspace}
\usepackage{amstext}
\usepackage{amsmath}
\usepackage{amssymb}
\usepackage{graphicx}

\usepackage{natbib}
\usepackage{aas_macros}
\usepackage{times}

\usepackage{ulem}

\newcommand{\FigDir}[1]{#1}

\def \kms {\ensuremath{\rm \,km\,s^{-1}}\xspace}

\def \kpc {\ensuremath{\rm {kpc}}\xspace}

\def \masyr {\ensuremath{\rm{mas}\,\rm{yr}^{-1}}\xspace}

\def \IVmI {\ensuremath{I_{\rm V-I}}\xspace}

\def \deg {\ensuremath{^{\circ}}}

\newlength{\voff}
\newcommand{\ppm}{$\pm$}

\begin{document}

\title[Proper Motion Dispersions of Red Clump Giants in the Galactic Bulge]{Proper Motion Dispersions of Red Clump Giants
  in the Galactic Bulge: Observations and Model Comparisons}

\author[Rattenbury et al.]
{Nicholas J. Rattenbury$^{1}$, Shude Mao$^{1}$,  Victor P. Debattista$^{2}$, Takahiro Sumi$^{3}$
\newauthor  Ortwin Gerhard$^{4}$, Flavio De Lorenzi$^4$ \thanks{e-mail: (njr, smao)@jb.man.ac.uk;  debattis@astro.washington.edu;  gerhard@exgal.mpe.mpg.de; sumi@stelab.nagoya-u.ac.jp; lorenzi@exgal.mpe.mpg.de}\\
$^1$ University of Manchester, Jodrell Bank Observatory, Macclesfield, Cheshire, SK11 9DL, UK \\
$^2$ Astronomy Department, University of Washington, Box 351580, Seattle, WA
98195-1580, USA \\
$^3$ Solar-Terrestrial Environment Laboratory, Nagoya University, Furo-cho, Chikusa-ku, Nagoya, 464-8601, Japan \\
$^4$ Max-Planck-Institut fuer extraterrestrische Physik, P.O. Box 1312,D-85741 Garching, Germany
}
\date{Accepted ........
      Received .......;
      in original form ......}

\pubyear{2005}

\maketitle
\begin{abstract}
Red clump giants in the Galactic bulge are approximate standard
candles and hence they can be used as distance indicators.  We compute
the proper motion dispersions of RCG stars in the Galactic bulge using the proper
motion catalogue from the second phase of the Optical Gravitational
Microlensing Experiment (OGLE-II, \citealt{2004MNRAS.348.1439S}) for
45 fields. The proper motion dispersions are measured to a few per
cent accuracy due to the large number of stars in the fields. The 
observational sample is comprised of 577736 stars.  These
observed data are compared to a state-of-the-art particle simulation
 of the Galactic bulge region. 
The predictions are in rough agreement with observations, but 
appear to be too anisotropic in the velocity ellipsoid. We note that
there is significant field-to-field variation in the observed proper motion dispersions. 
This could either be a real feature, or due to some unknown systematic effect. 

\end{abstract}

\begin{keywords}
gravitational lensing - Galaxy: bulge - Galaxy: centre - Galaxy:
kinematics and dynamics - Galaxy: structure
\end{keywords}

\section{Introduction}
\label{sec:intro}

Many lines of evidence suggest the presence of a bar at the Galactic
centre, such as infrared maps \citep{1995ApJ...445..716D,1997MNRAS.288..365B} and star
counts
\citep{1997ApJ...477..163S,1997ApJ...487..885N,1998MNRAS.295..145U},
see \citet{2002ASPC..273...73G} for a review. However, the bar
parameters are not well determined. For example, recent infra-red star
counts collected by the Spitzer Space Telescope are best explained
assuming a bar at a $\sim 44\deg$ angle to the Sun--Galactic centre
line \citep{2005ApJ...630L.149B} while most previous studies prefer a
bar at $\sim 20\deg$. In addition, there may be some fine features,
such as a ring in the Galactic bulge, that are not yet firmly
established \citep{2005MNRAS.358.1309B}. It is therefore crucial to
obtain as many constraints as possible in order to better understand
the structure of the inner Galaxy.

Many microlensing groups monitor the Galactic bulge, including the
EROS \citep{1993Natur.365..623A}, MACHO \citep{2000ApJ...541..734A},
MOA \citep{2001MNRAS.327..868B, 2003ApJ...591..204S} and OGLE
\citep{2000AcA....50....1U} collaborations. In addition to discovering
microlensing events, these groups have also accumulated a huge amount
of data about the stars in the Galactic bulge spanning several years
to a decade and a half.

\citet{2001MNRAS.327..601E} first demonstrated that the data can be
used to infer the proper motions of stars, down to $\sim
\masyr$. \citet{2004MNRAS.348.1439S}
obtained the
proper motions for millions of stars in the OGLE-II database for a
large area of the sky. In this paper, we focus on the red clump
giants. These stars are bright and they are approximately standard
candles, hence their magnitudes can be taken as a crude measure of
their distances. As the OGLE-II proper motions are relative, in this
paper we compute the proper motion dispersions of bulge stars for all field data
presented by \citet{2004MNRAS.348.1439S}, as they are independent of
the unknown proper motion zero-points. These results could aid
theoretical modelling efforts for the central regions of the Galaxy.

The structure of the paper is as follows. In section~\ref{sec:OGLE}, we describe the OGLE-II proper motion catalogue and compute the proper motion dispersions for bulge stars in 45 OGLE-II fields. In section~\ref{sec:model} we describe the stellar-dynamical model of the Galaxy used in this work and detail how the model was used to generate proper motion dispersions. These model predictions are compared to the observational results in section~\ref{sec:results} and in section~\ref{sec:discussion} we discuss the implications of the results.

\section{Observed Proper Motion Dispersions}
\label{sec:OGLE}
\begin{figure*}
\psfrag{xlabel}{\hspace{-10pt}\normalsize Galactic longitude $(\deg)$}
\psfrag{ylabel}{\hspace{-20pt}\normalsize Galactic latitude $(\deg)$}
\psfrag{L}{\raisebox{5pt}{\scriptsize l}}
\psfrag{B}{\raisebox{-5pt}{\scriptsize b}}

\centering \includegraphics[height=0.98\hsize, angle=-90]{\FigDir{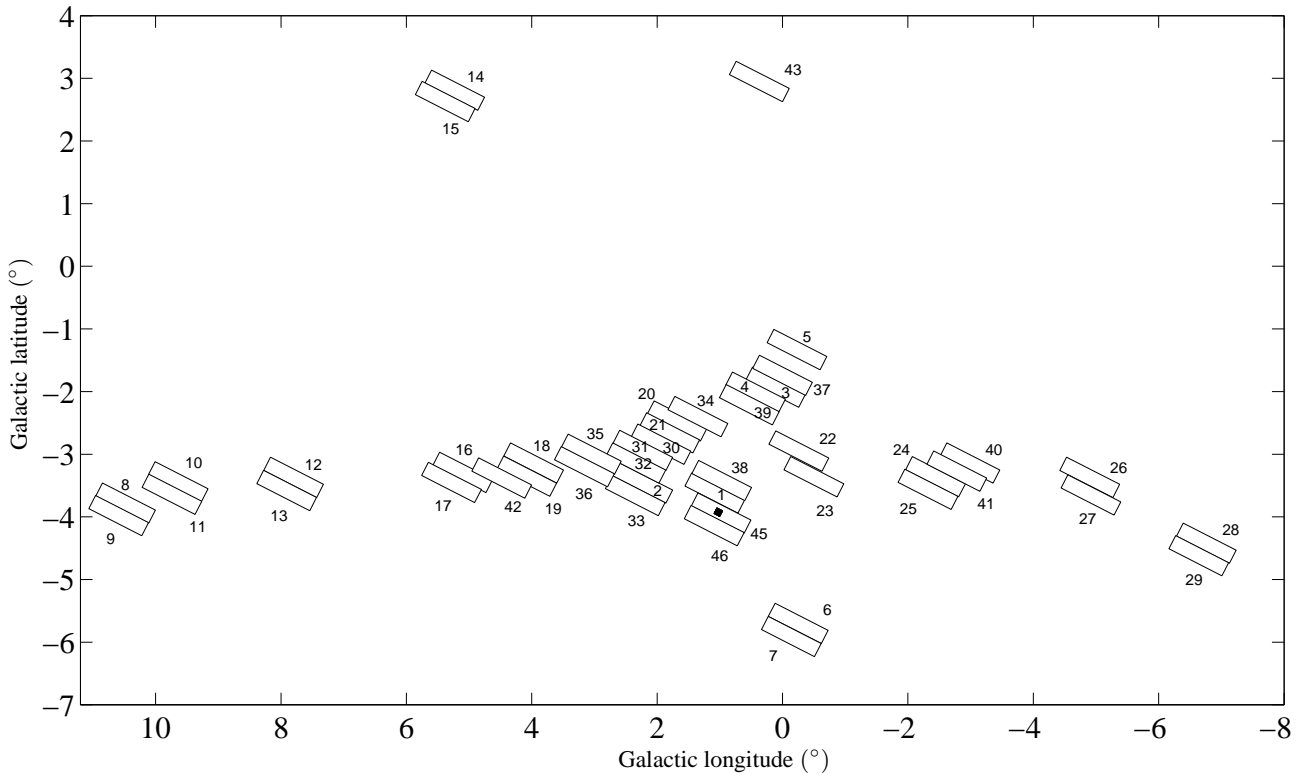}}
\caption{The position of the 45 OGLE-II fields used in this analysis. The field used in \citet{1992AJ....103..297S} is shown, located within OGLE-II field 45 with $(l,b)=(1.0245\deg, -3.9253\deg)$.}
\label{fig:fields}
\end{figure*}

The second phase of the OGLE experiment observed the Galactic Centre in 49 fields using the 1.3m Warsaw telescope at the Las Campanas Observatory, Chile. Data were collected over an interval of almost four years, between 1997 and 2000. Each field is $0.24\deg \times 0.95\deg$ in size. Fig.~\ref{fig:fields} shows the position of the OGLE-II Galactic Bulge fields which returned data used in this paper.

\subsection{Red Clump Giants}

The red clump giants are metal-rich horizontal branch stars \citetext{\citealt{2000AcA....50..191S} and references therein}. Theoretically, one expects their magnitudes to have (small) variations with metallicity, age and initial stellar mass \citep{2001MNRAS.323..109G}. Empirically they appear to be reasonable standard candles in the $I$-band with little dependence on metallicities \citep{2000ApJ...531L..25U, 2001ApJ...551L..85Z}. Below we describe the selection of RCG stars in more detail.

\subsection{OGLE-II proper motion data}
\label{sec:observed}
\begin{figure}
\psfrag{ylabel}{\hspace{0pt}{\normalsize $I_{0}$}}
\psfrag{xlabel}{\hspace{0pt}{\normalsize $(V-I)_{0}$}}
\begin{center}
\hspace{-1cm}
\centering\includegraphics[width=1.0\hsize]{\FigDir{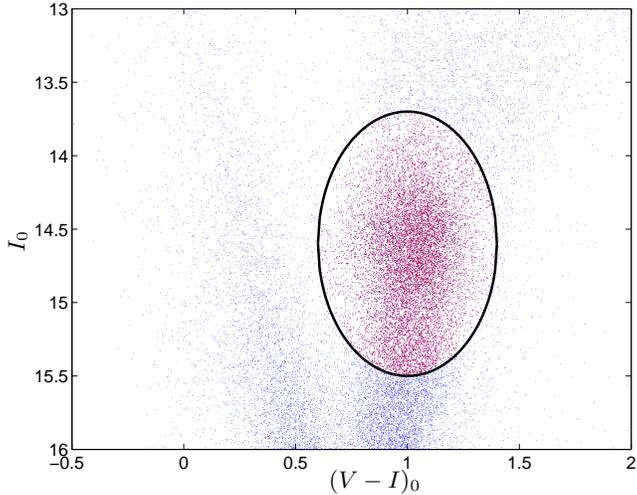}}
\end{center}
\caption{Extinction-corrected colour-magnitude diagram for stars in the OGLE-II field 1. The ellipse defines the selection criteria for RCG stars based on colour and magnitude, see text. Sample stars are also required to  have proper motion errors $s_{\rm l,b} < 1$ \masyr and total proper motion $\mu < 10$ \masyr.}
\label{fig:cmd}

\end{figure}

Bulge RCG stars are selected from the OGLE-II proper motion catalogue
by applying a cut in magnitude and colour to all stars in each of the
OGLE-II fields. We corrected for extinction and reddening using the
maps presented by \citet{2004MNRAS.349..193S} for each field. Stars
were selected which are located in an ellipse with centre $(V-I)_{0} =
1.0$ , $I_{0} = 14.6$; and semi-major (magnitude) and semi-minor
(colour) axes of 0.9 and 0.4 respectively, see Fig.~\ref{fig:cmd}; a
similar selection criterion was used by Sumi (2004). Stars with errors
in proper motion greater than 1~\masyr in either the $l$ or $b$
directions were excluded. Stars with total proper motion greater than
10 \masyr where similarly excluded, as these are likely to be nearby
disk stars, see also section~\ref{sec:modkine}. Fields 44, 47-49 were not analysed due to the low number 
of RCG stars appearing in these fields. 

The proper motion dispersions for the longitude and latitude directions ($\sigma_{\rm l}$ and $\sigma_{\rm b}$) were computed for each field via a maximum likelihood analysis following \citet{1987AJ.....93.1114L}. Assuming a Gaussian distribution of proper motions with mean $\bar{\mu}$ and intrinsic proper motion dispersion $\sigma$, the probability of a single observed proper motion $\mu_{i}$ with measurement error $\xi_{i}$ is:
\begin{equation}
p_{i} = \frac{1}{\sqrt{2\pi (\sigma^{2} + \xi_{i}^{2})}} \exp\left[-\frac{(\mu_{i} - \bar{\mu})^{2}}{2(\sigma^{2} + \xi_{i}^{2})}\right]
\end{equation}
Maximising the likelihood $\ln(L) = \ln(\prod p_{i})$ for $\bar{\mu}$ and $\sigma$ over all observations we find:
\begin{equation}
\frac{\partial \ln L}{\partial \bar{\mu}} = \sum_{i} \frac{(\mu_{i} - \bar{\mu})}{\sigma^{2} + \xi_{i}^{2}} = 0
\end{equation}
\begin{equation}
\Rightarrow \bar{\mu} = \sum_{i} \frac{\mu_{i}}{\sigma^{2} + \xi_{i}^{2}} \Bigg{\slash} \sum_{i} (\sigma^{2} + \xi_{i}^{2})^{-1}
\end{equation}
and
\begin{equation}
\frac{\partial \ln L}{\partial \sigma} = \sum_{i} \frac{1}{\sigma^{2} + \xi_{i}^{2}} - \sum_{i} \frac{(\mu_{i} - \bar{\mu})^{2}}{(\sigma^{2} + \xi_{i}^{2})^{2}} = 0
\end{equation}
which can be solved numerically to find $\sigma^{2}$.

The values of $\bar{\mu}$ and $\sigma$ obtained using the above maximum-likelihood analysis are virtually identical to those obtained via the equations in \citet{1992AJ....103..297S}. The errors on the observed proper motion dispersion values were determined from a bootstrap analysis using 500 samplings of the observed dataset.

\subsection{Extinction}
In order to ensure the correction for extinction and reddening above does not affect the kinematic measurements,  $\sigma_{\rm l}$ and  $\sigma_{\rm b}$ were recomputed for each OGLE-II field using reddening-independent magnitudes. Following \citet{1997ApJ...477..163S} we define the reddening-independent magnitude $I_{\rm V-I}$:
\begin{equation}
\IVmI = I - A_{\rm I} / (A_{\rm V} - A_{\rm I})\; (V-I)
\end{equation}
where $A_{\rm I}$ and $A_{\rm V}$ are the extinctions in the $I$ and $V$ bands determined by \citet{2004MNRAS.349..193S}. The position of the red clump in the $I_{\rm V-I}$, $(V-I)$ CMD varies from field to field. The red clump stars were extracted by iteratively applying a selection ellipse computed from the moments of the data \citep{Roc02} rather than centred on a fixed colour and magnitude. The selection ellipse was recomputed iteratively for each sample until convergence. The proper motion dispersions $\sigma_{\rm l}$ and  $\sigma_{\rm b}$ computed using RCG stars selected in this way are consistent with those determined using the original selection criteria on corrected magnitudes and colours.

\subsection{Results}

Table~\ref{tab:obsresults} lists the observed proper motion dispersions along with errors for each of the 45 OGLE-II fields considered in this paper. 

\begin{figure}
\psfrag{xlabel}{\normalsize \raisebox{-2pt}{\hspace{-25pt} Galactic longitude $(\deg)$}}
\psfrag{ylabel1}{\normalsize \hspace{-10pt} $\sigma_{\rm l}$ (\masyr)}
\psfrag{ylabel2}{\normalsize \hspace{-10pt} $\sigma_{\rm b}$ (\masyr)}

\centering\includegraphics[width=1.0\hsize]{\FigDir{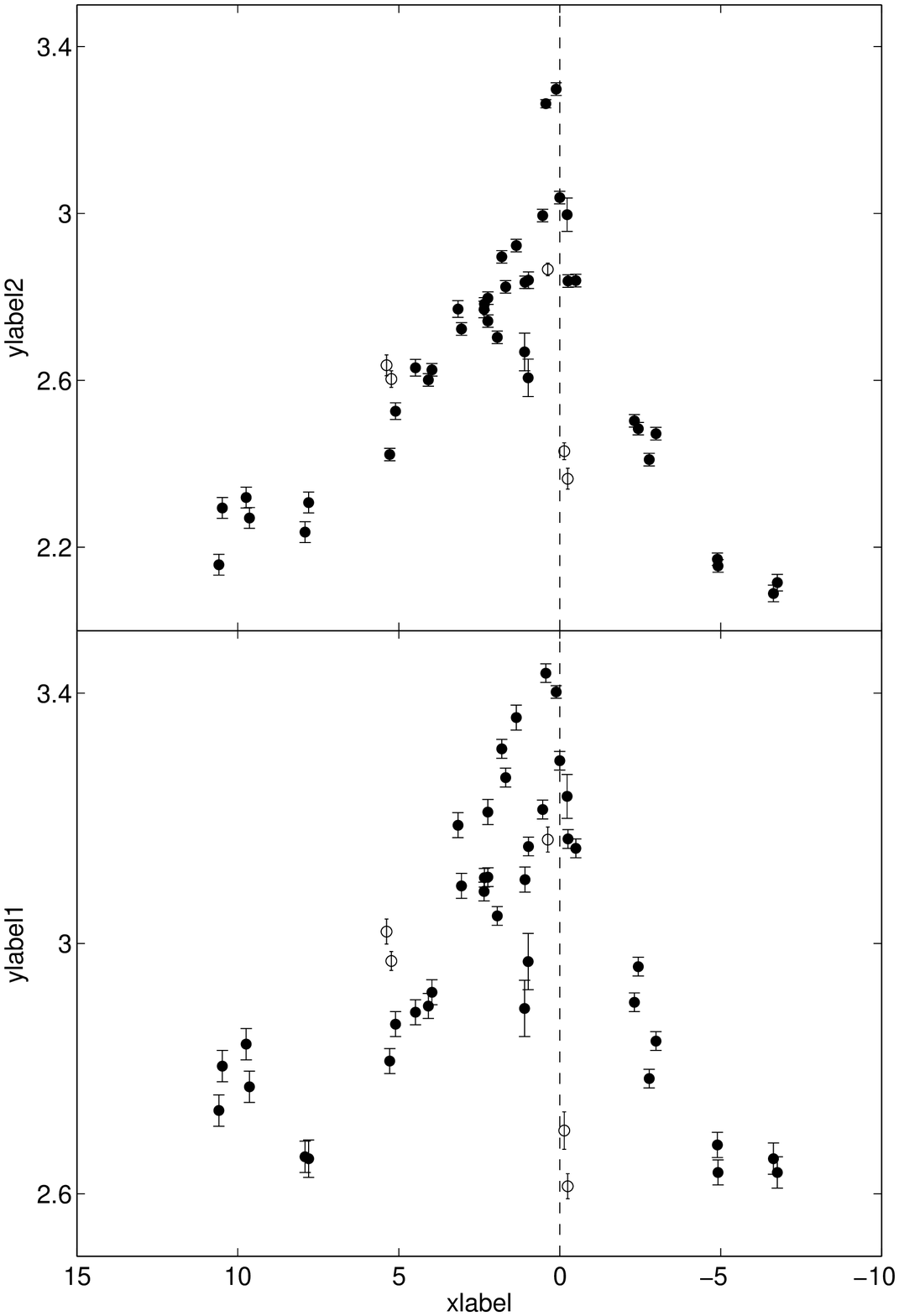}}
\caption{\label{fig:sigmas_vs_L}Proper motion dispersion in the Galactic longitude ($\sigma_{\rm l}$) and latitude ($\sigma_{\rm b}$) directions for 45 OGLE-II Galactic bulge fields as a function of field Galactic longitude. Open circles correspond to fields 6, 7, 14, 15 and 43 which have relatively extreme galactic latitudes, see Fig.~\ref{fig:fields}.}

\end{figure}

\begin{figure}
\psfrag{xlabel}{\normalsize \raisebox{-2pt}{\hspace{-25pt} Galactic latitude $(\deg)$}}
\psfrag{ylabel1}{\normalsize \hspace{-10pt} $\sigma_{\rm b}$ (\masyr)}
\psfrag{ylabel2}{\normalsize \hspace{-10pt} $\sigma_{\rm l}$ (\masyr)}

\centering\includegraphics[width=1.0\hsize]{\FigDir{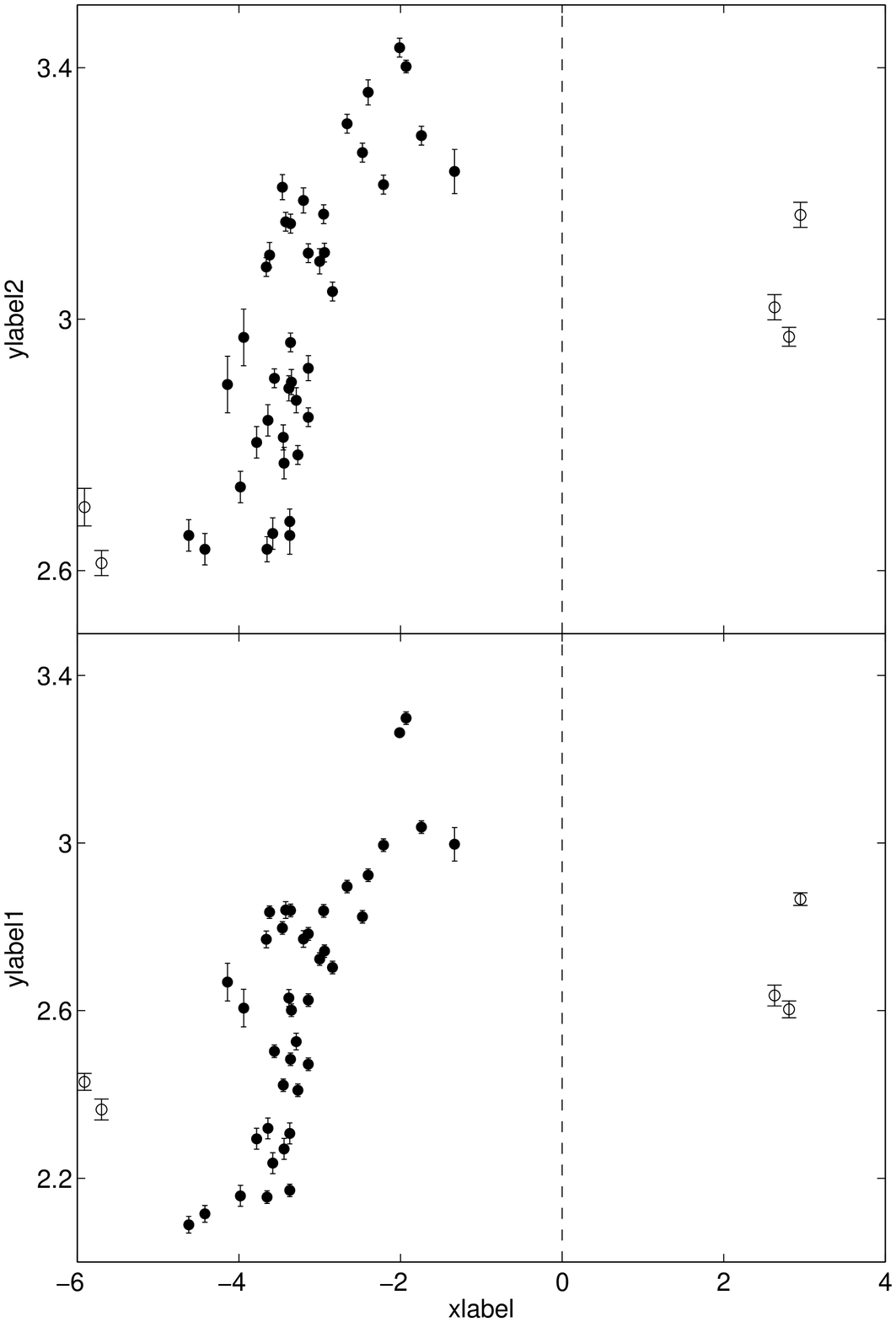}}
\caption{\label{fig:sigmas_vs_B}Proper motion dispersion in the Galactic longitude ($\sigma_{\rm l}$) and latitude ($\sigma_{\rm b}$) directions for 45 OGLE-II Galactic bulge fields as a function of field Galactic latitude. Open circles correspond to fields 6, 7, 14, 15 and 43 which have relatively extreme galactic latitudes, see Fig.~\ref{fig:fields}.}
\end{figure}

Figures~\ref{fig:sigmas_vs_L} and \ref{fig:sigmas_vs_B} show the proper motion dispersions $\sigma_{\rm l}$ and  $\sigma_{\rm b}$ as a function of Galactic longitude and latitude.  A typical value of $\sigma_{\rm l}$ or $\sigma_{\rm b}$ of 3.0 \masyr corresponds to $\sim110$ \kms, assuming a distance to the Galactic centre of 8 \kpc. The proper motion dispersion profiles as a function of Galactic longitude shows some slight asymmetry about the Galactic centre. This asymmetry may be related to the tri-axial Galactic bar structure \citep{1997ApJ...477..163S,2005ApJ...621L.105N,2005MNRAS.358.1309B}. The most discrepant points in Fig.~\ref{fig:sigmas_vs_L} correspond to the low-latitude fields numbers 6 and 7 (see Fig.~\ref{fig:fields}). The varying field latitude accounts for some of the scatter in Fig.~\ref{fig:sigmas_vs_L}, however we note below in section~\ref{sec:diff} that there are significant variations in the observed proper motion dispersion between some pairs of adjacent fields.  Owing to the the lack of fields at positive Galactic latitude, any asymmetry about the Galactic centre in the proper motion dispersions as a function of Galactic latitude is not obvious, see Fig.~\ref{fig:sigmas_vs_B}. Field-to-field variations in the proper motion dispersions similarly contribute to the scatter seen in Fig.~\ref{fig:sigmas_vs_B}, along with the wide range of field longitudes, especially for fields with $-4\deg < b < -3\deg$. 

Table~\ref{tab:compKoz} lists the proper motion dispersions and cross-correlation term $C_{\rm lb}$ in the OGLE-II Baade's Window fields 45 and 46 along with those found by \citet{2006MNRAS.370..435K} using HST data in four BW fields. The two sets of proper motion dispersions results are consistent at the $\sim 2\sigma$ level. It is important to note that the errors on the proper motion dispersions in Table~\ref{tab:obsresults} do not include systematic errors. We also note that the selection criteria applied to stars in the HST data are very different to those for the ground-based data, in particular the magnitude limits applied in each case. The bulge kinematics from the HST data of \citet{2006MNRAS.370..435K} were determined for stars with magnitudes $18.0 < I_{\rm F814W} < 21.5$. The approximate reddening-independent magnitude range for the OGLE-II data was $ 12.5 \lesssim \IVmI \lesssim 14.6 $.  The effects of blending are also very different in the two datasets. It is therefore very reassuring that our results are in general agreement with those obtained by \citet{2006MNRAS.370..435K} using higher resolution data from the HST. For more comparisons between ground and HST RCG proper motion dispersions, see section~\ref{sec:results}.

\begin{figure}
\psfrag{xlabel1}{\normalsize \raisebox{-2pt}{\hspace{-25pt} Galactic longitude $(\deg)$}}
\psfrag{xlabel2}{\normalsize \raisebox{-2pt}{\hspace{-25pt} Galactic latitude $(\deg)$}}
\psfrag{ylabel1}{\normalsize \hspace{0pt} $C_{\rm lb}$}
\psfrag{ylabel2}{\normalsize \hspace{0pt} $C_{\rm lb}$}

\centering\includegraphics[width=1.0\hsize]{\FigDir{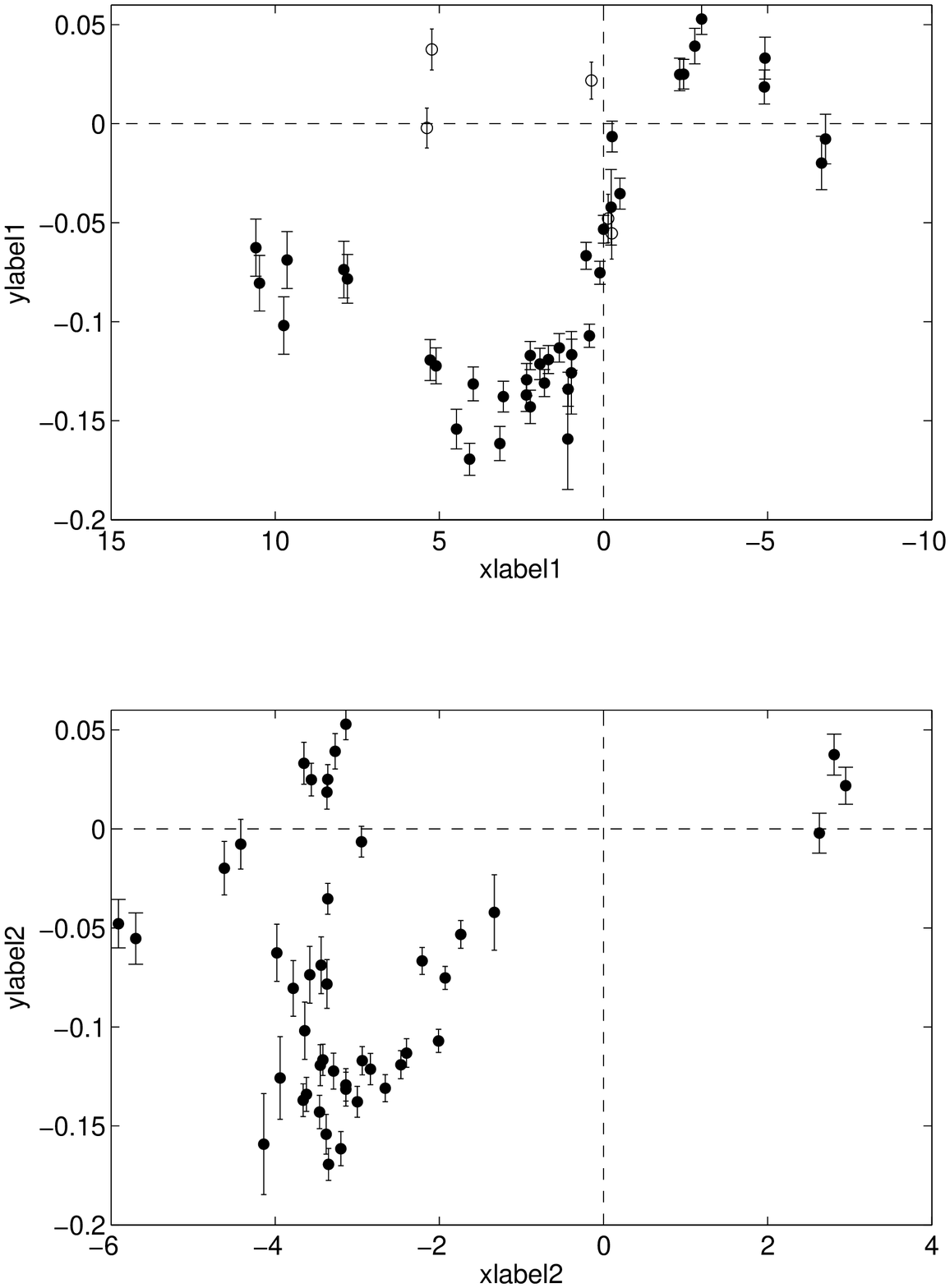}}
\caption{\label{fig:Clb_vs_LandB}Cross-correlation term $C_{\rm lb}$ for 45 OGLE-II Galactic bulge fields as a function of field Galactic longitude (top) and latitude (bottom). Open circles in the top plot of $C_{\rm lb}$ vs. $l$ correspond to fields 6, 7, 14, 15 and 43 which have relatively extreme galactic latitudes, see Fig.~\ref{fig:fields}.}

\end{figure}

Figure~\ref{fig:Clb_vs_LandB} shows the cross-correlation term $C_{\rm lb}$ as a function of Galactic co-ordinate. There is a clear sinusoidal structure in the $C_{\rm lb}$ data as a function of Galactic longitude, with the degree of correlation between $\sigma_{\rm l}$ and  $\sigma_{\rm b}$ changing most rapidly near $\l\simeq0\deg$. The $C_{\rm lb}$ data as a function of Galactic latitude may also show some evidence of structure. It is possible however, that this apparent structure is due to the different number of fields at each latitude, rather than some real physical cause.

\begin{table*}
\caption{\label{tab:obsresults}Observed proper motion dispersions in the longitude and latitude directions, $\sigma_{\rm l}$, $\sigma_{\rm b}$ , and cross-correlation term $C_{\rm lb}$ for bulge stars in 45 OGLE-II fields. High precision proper motion data for bulge stars were
extracted from the OGLE-II proper motion catalogue
\citep{2004MNRAS.348.1439S}.  $N$ is the number of stars selected from each field. Fields 44, 47-49 were not analysed due to the low number  of RCG stars appearing in these fields. }
  \begin{tabular}{crrcccrr}
\hline 
Field & \multicolumn{2}{c}{Field centre}  &  \multicolumn{2}{c}{PM Dispersions (\masyr)} &  $C_{\rm lb}$ & $N$\phantom{x} \\
& $l\,(\deg)$ & $b\,(\deg)$  & Longitude $\sigma_{\rm l}$ & Latitude $\sigma_{\rm b}$ & & \\

\hline
1 & 1.08 & -3.62 & 3.10  \ppm  0.02 & 2.83  \ppm  0.02 & -0.13  \ppm  0.01  & 15434 \\ 
2 & 2.23 & -3.46 & 3.21  \ppm  0.02 & 2.80  \ppm  0.02 & -0.14  \ppm  0.01  & 16770 \\ 
3 & 0.11 & -1.93 & 3.40  \ppm  0.01 & 3.30  \ppm  0.02 & -0.08  \ppm  0.01  & 26763 \\ 
4 & 0.43 & -2.01 & 3.43  \ppm  0.02 & 3.26  \ppm  0.01 & -0.11  \ppm  0.01  & 26382 \\ 
5 & -0.23 & -1.33 & 3.23  \ppm  0.03 & 3.00  \ppm  0.04 & -0.04  \ppm  0.02  & 3145 \\ 
6 & -0.25 & -5.70 & 2.61  \ppm  0.02 & 2.36  \ppm  0.03 & -0.06  \ppm  0.01  & 7027 \\ 
7 & -0.14 & -5.91 & 2.70  \ppm  0.03 & 2.43  \ppm  0.02 & -0.05  \ppm  0.01  & 6236 \\ 
8 & 10.48 & -3.78 & 2.80  \ppm  0.03 & 2.29  \ppm  0.02 & -0.08  \ppm  0.01  & 5136 \\ 
9 & 10.59 & -3.98 & 2.73  \ppm  0.02 & 2.16  \ppm  0.03 & -0.06  \ppm  0.01  & 5114 \\ 
10 & 9.64 & -3.44 & 2.77  \ppm  0.02 & 2.27  \ppm  0.02 & -0.07  \ppm  0.01  & 5568 \\ 
11 & 9.74 & -3.64 & 2.84  \ppm  0.02 & 2.32  \ppm  0.02 & -0.10  \ppm  0.01  & 5369 \\ 
12 & 7.80 & -3.37 & 2.66  \ppm  0.03 & 2.31  \ppm  0.03 & -0.08  \ppm  0.01  & 6035 \\ 
13 & 7.91 & -3.58 & 2.66  \ppm  0.03 & 2.24  \ppm  0.02 & -0.07  \ppm  0.01  & 5601 \\ 
14 & 5.23 & 2.81 & 2.97  \ppm  0.02 & 2.60  \ppm  0.02 & 0.04  \ppm  0.01  & 10427 \\ 
15 & 5.38 & 2.63 & 3.02  \ppm  0.02 & 2.64  \ppm  0.03 & -0.00  \ppm  0.01  & 8989 \\ 
16 & 5.10 & -3.29 & 2.87  \ppm  0.02 & 2.53  \ppm  0.02 & -0.12  \ppm  0.01  & 9799 \\ 
17 & 5.28 & -3.45 & 2.81  \ppm  0.02 & 2.42  \ppm  0.01 & -0.12  \ppm  0.01  & 10268 \\ 
18 & 3.97 & -3.14 & 2.92  \ppm  0.02 & 2.62  \ppm  0.02 & -0.13  \ppm  0.01  & 14019 \\ 
19 & 4.08 & -3.35 & 2.90  \ppm  0.02 & 2.60  \ppm  0.02 & -0.17  \ppm  0.01  & 13256 \\ 
20 & 1.68 & -2.47 & 3.27  \ppm  0.01 & 2.82  \ppm  0.01 & -0.12  \ppm  0.01  & 17678 \\ 
21 & 1.80 & -2.66 & 3.31  \ppm  0.02 & 2.90  \ppm  0.02 & -0.13  \ppm  0.01  & 17577 \\ 
22 & -0.26 & -2.95 & 3.17  \ppm  0.02 & 2.84  \ppm  0.02 & -0.01  \ppm  0.01  & 19787 \\ 
23 & -0.50 & -3.36 & 3.15  \ppm  0.01 & 2.84  \ppm  0.02 & -0.04  \ppm  0.01  & 17996 \\ 
24 & -2.44 & -3.36 & 2.96  \ppm  0.01 & 2.48  \ppm  0.01 & 0.02  \ppm  0.01  & 16397 \\ 
25 & -2.32 & -3.56 & 2.91  \ppm  0.01 & 2.50  \ppm  0.01 & 0.02  \ppm  0.01  & 16386 \\ 
26 & -4.90 & -3.37 & 2.68  \ppm  0.02 & 2.17  \ppm  0.01 & 0.02  \ppm  0.01  & 13099 \\ 
27 & -4.92 & -3.65 & 2.63  \ppm  0.02 & 2.15  \ppm  0.01 & 0.03  \ppm  0.01  & 12728 \\ 
28 & -6.76 & -4.42 & 2.63  \ppm  0.03 & 2.12  \ppm  0.02 & -0.01  \ppm  0.01  & 8367 \\ 
29 & -6.64 & -4.62 & 2.66  \ppm  0.03 & 2.09  \ppm  0.02 & -0.02  \ppm  0.01  & 8108 \\ 
30 & 1.94 & -2.84 & 3.04  \ppm  0.02 & 2.70  \ppm  0.02 & -0.12  \ppm  0.01  & 17774 \\ 
31 & 2.23 & -2.94 & 3.11  \ppm  0.02 & 2.74  \ppm  0.01 & -0.12  \ppm  0.01  & 17273 \\ 
32 & 2.34 & -3.14 & 3.10  \ppm  0.02 & 2.78  \ppm  0.01 & -0.13  \ppm  0.01  & 15966 \\ 
33 & 2.35 & -3.66 & 3.08  \ppm  0.02 & 2.77  \ppm  0.02 & -0.14  \ppm  0.01  & 15450 \\ 
34 & 1.35 & -2.40 & 3.36  \ppm  0.02 & 2.92  \ppm  0.01 & -0.11  \ppm  0.01  & 16889 \\ 
35 & 3.05 & -3.00 & 3.09  \ppm  0.02 & 2.72  \ppm  0.02 & -0.14  \ppm  0.01  & 15973 \\ 
36 & 3.16 & -3.20 & 3.19  \ppm  0.02 & 2.77  \ppm  0.02 & -0.16  \ppm  0.01  & 14955 \\ 
37 & 0.00 & -1.74 & 3.29  \ppm  0.02 & 3.04  \ppm  0.01 & -0.05  \ppm  0.01  & 20233 \\ 
38 & 0.97 & -3.42 & 3.15  \ppm  0.01 & 2.84  \ppm  0.02 & -0.12  \ppm  0.01  & 15542 \\ 
39 & 0.53 & -2.21 & 3.21  \ppm  0.01 & 3.00  \ppm  0.01 & -0.07  \ppm  0.01  & 24820 \\ 
40 & -2.99 & -3.14 & 2.84  \ppm  0.01 & 2.47  \ppm  0.02 & 0.05  \ppm  0.01  & 13581 \\ 
41 & -2.78 & -3.27 & 2.78  \ppm  0.01 & 2.41  \ppm  0.02 & 0.04  \ppm  0.01  & 14070 \\ 
42 & 4.48 & -3.38 & 2.89  \ppm  0.02 & 2.63  \ppm  0.02 & -0.15  \ppm  0.01  & 10099 \\ 
43 & 0.37 &  2.95 & 3.17  \ppm  0.02 & 2.87  \ppm  0.01 &  0.02  \ppm  0.01  & 11467 \\ 
45 & 0.98 & -3.94 & 2.97  \ppm  0.04 & 2.61  \ppm  0.04 & -0.13  \ppm  0.02  & 2380 \\ 
46 & 1.09 & -4.14 & 2.90  \ppm  0.04 & 2.67  \ppm  0.04 & -0.16  \ppm  0.03  & 1803 \\ 

\hline
\end{tabular}

\end{table*}

\begin{table*}
\caption{\label{tab:compKoz}Comparison between proper motion dispersions and cross-correlation term $C_{\rm lb}$ in two of the OGLE-II fields (45 and 46) with proper motion dispersions computed from four nearby HST fields \citep{2006MNRAS.370..435K}.}
\begin{tabular}{ccccccc}
\hline
Field & $l\,(\deg)$ & $b\,(\deg)$ & $\sigma_{\rm l}$ (\masyr)&  $\sigma_{\rm b}$ (\masyr)& $C_{\rm lb}$ & Ref \\ \hline
119-A &      1.32 & -3.77 & 2.89 \ppm 0.10 & 2.44 \ppm 0.08 & -0.14 \ppm 0.04 & $^{1}$\\ 
119-C &      0.85 & -3.89 & 2.79 \ppm 0.10 & 2.65 \ppm 0.08 & -0.14 \ppm 0.04 & $^{1}$\\ 
OGLE-II 45 &         0.98 & -3.94 & 2.97 \ppm 0.04 & 2.61 \ppm 0.04 & -0.13 \ppm 0.02 & $^{2}$ \\[10pt] 
119-D &      1.06 & -4.12 & 2.75 \ppm 0.10 & 2.56 \ppm 0.09 & -0.05 \ppm 0.06 & $^{1}$ \\ 
95-BLG-11 &  0.99 & -4.21 & 2.82 \ppm 0.09 & 2.62 \ppm 0.09 & -0.14 \ppm 0.04 & $^{1}$ \\ 
OGLE-II 46 &         1.09 & -4.14 & 2.90 \ppm 0.04 & 2.67 \ppm 0.04 & -0.16 \ppm 0.03 & $^{2}$\\ 
\hline
\multicolumn{3}{l}{$^{1}$\citet{2006MNRAS.370..435K}  $^{2}$This work. }
\end{tabular}

\end{table*}

\section{Galactic Model}
\label{sec:model}
The stellar-dynamical model used in this work was produced using the
made-to-measure method \citep{1996MNRAS.282..223S}. The model is
constrained to reproduce the density distribution constructed from the
dust-corrected $L$-band  COBE/DIRBE map of \citet{1996sgni.conf..128S}.  An earlier dynamical model was built to match the total column density of the disk \citep{2002MNRAS.330..591B}.  This dynamical model matched the radial
velocity and proper motion data in two fields (including Baade's
window) quite well. No kinematic constraints were imposed during the construction of the model. We refer the readers to
\citet{2004ApJ...601L.155B} for more detailed descriptions.
The model used here is constructed as in that case with the further
refinement that the vertical density distribution is also included.
This is necessary as the vertical kinematics ($\sigma_{\rm b}$) will also be
compared with observations in this paper.  However the density
distribution near the mid-plane is considerably more uncertain, in
part because of the dust extinction correction.  Thus the model used
in this paper can only be considered illustrative, not final.  Further
efforts to model the vertical density distribution are currently under
way and will be reported elsewhere (Debattista et al. 2007, in preparation).

\begin{figure}
\hspace{0.0cm}
\includegraphics[width=1.0\hsize]{\FigDir{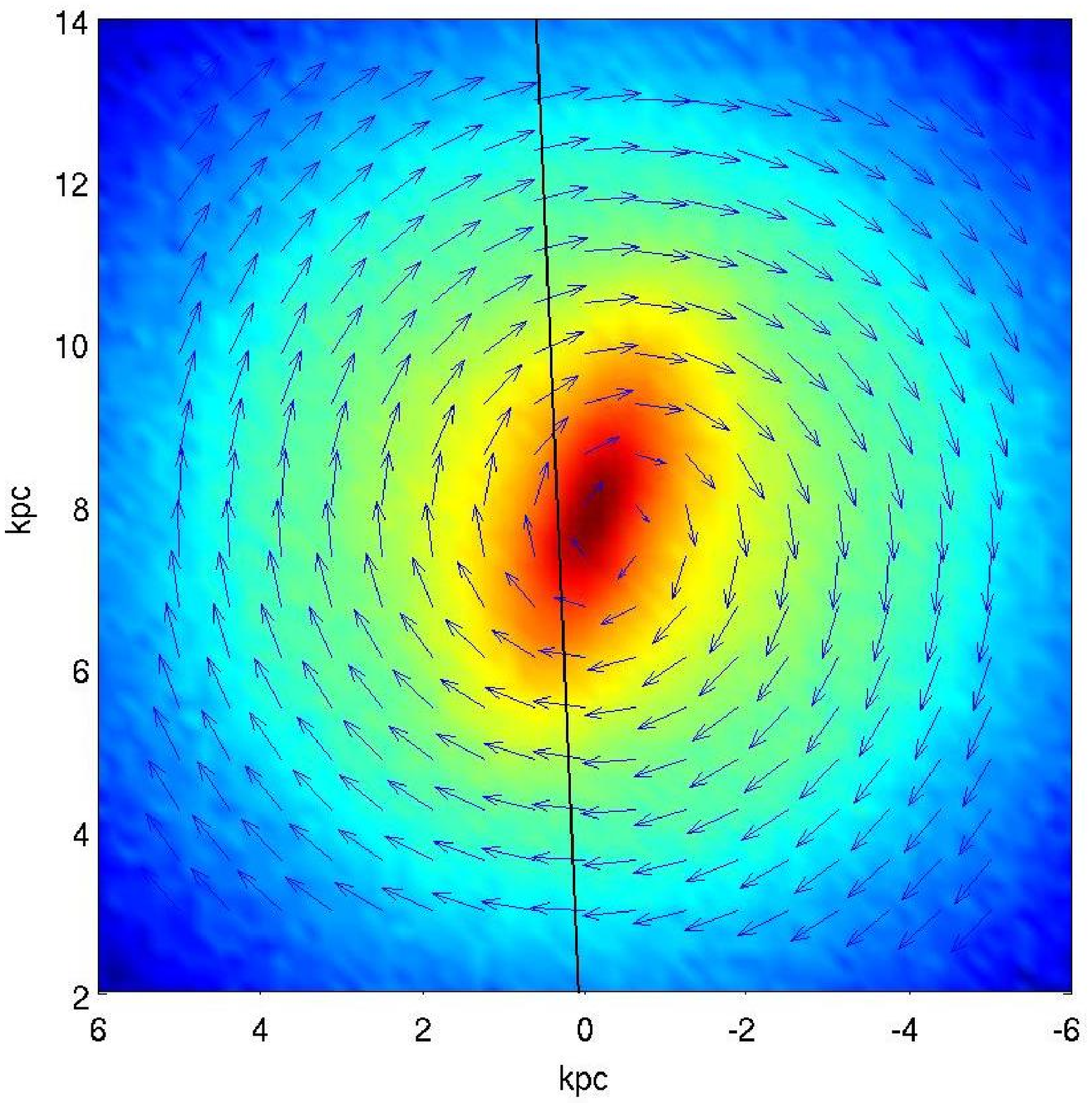}}
\caption{
Galactic kinematics from the model of Debattista et~al (2007, in preparation). Bulk stellar motion in the mid-plane of the Galaxy is shown super-imposed on the stellar density. The Sun is located at the origin (not shown). An example line of sight is shown. The model can be rotated to four equivalent positions for each line of sight due to symmetry (see section~\ref{sec:modelmags}).}
\label{fig:velocity}
\end{figure}

In Fig. \ref{fig:velocity}, we present the mean motion of stars in the
mid-plane of the Galaxy from this model.  A bar position angle of
$\theta = 20\deg$ is shown here, as this is the orientation
favoured both by optical depth measurements \citep{2002ApJ...567L.119E} and by the red clump giant brightness distribution \citep{1997ApJ...477..163S} and
was the angle used in deriving the model. Clearly one can see that the mean motion follows elliptical paths around the Galactic bar. The analysis of OGLE-II proper motions by \citet{2003MNRAS.340.1346S} is consistent with this streaming motion.

\subsection{Model stellar magnitudes}
\label{sec:modelmags}
The model has a four-fold symmetry, obtained by a rotation of $\pi$ radians around the vertical axis and by positioning the Sun above or below the mid-plane. The kinematics of model particles falling within the solid angle of each OGLE-II field were combined to those from the three other equivalent lines-of-sight. This procedure allows an increase in the number of model particles used for the predictions of stellar kinematics. 

\begin{figure}
\psfrag{xlabel}{\normalsize $I$}
\psfrag{ylabel}{\normalsize $N_{1}(I)$}
\psfrag{embed1}{\hspace{10pt}\raisebox{30pt}{\normalsize \parbox{20pt}{RCG stars}}}
\psfrag{embed2}{\normalsize non-RCG stars}

\begin{center}
\hspace{-1cm}
\centering\includegraphics[width=1.0\hsize]{\FigDir{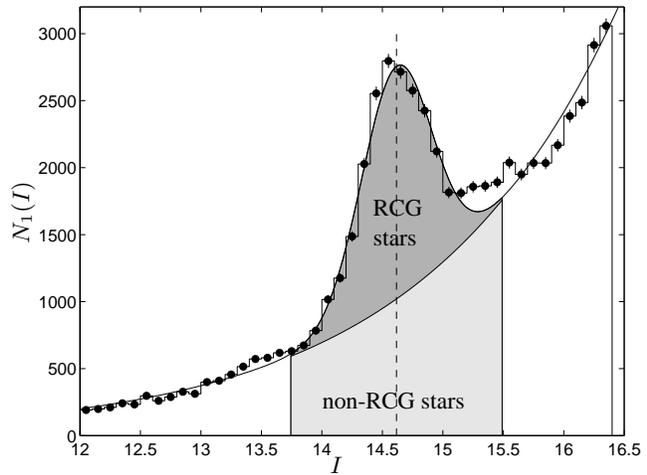}}
\end{center}
\caption{ Number count as function of apparent magnitude,
$I$, for OGLE-II field 1. The number count histogram is shown along with the fitted function equation~(\ref{eq:count}). The fraction of RCG stars, $R_{k}$, is evaluated
over the magnitude range $I_{\text{p}} \pm 3\sigma$ for each of the
($k=1\ldots49$) OGLE-II fields. The ratio $R_{k}$ is assumed to be the
same at all stellar distances for each field.}
\label{fig:ratioplot}
\end{figure}

We assign magnitudes to stars in the Galactic model described above
which appear in the same fields as that observed by the OGLE
collaboration. Number counts as a function of $I$-band apparent
magnitude, $I$, were used to compute the fraction of RCG
stars in each of the OGLE-II fields.  Figure~\ref{fig:ratioplot} shows
an example of the fitted number count function $N_{k}(I)$
for one of the $k=1\ldots49$ OGLE-II fields, where
$N_{k}(I)$ is of the form of a power-law and a Gaussian
\citep{2004MNRAS.349..193S}:
\begin{equation}
N_{k}(I) = a_{k}10^{(b_{k}I)} + c_{k}\exp \left[ \frac{ -(I-I_{\text{p},k})^2}{2\sigma_{k}^{2}} \right]
\label{eq:count}
\end{equation}
where the constants $ a_{k}, b_{k}, c_{k}, I_{\text{p},k}, \sigma_{k}$
are determined for each of the $k$ OGLE-II fields, see Table~\ref{tab:fieldconsts}.
The fraction $R_{k}$ of RCG stars is evaluated as the ratio of the area under the Gaussian component of
equation~(\ref{eq:count}) to the area under the full expression. The
integrals are taken over $\pm3\sigma_{k}$ around the RCG peak in
$N_{k}(I)$ for each of the $k$ OGLE-II fields.  Fields 44
and 47-49 are not included as there are insufficient RCGs in the
OGLE-II fields to fit equation~(\ref{eq:count}). Figure~\ref{fig:ratioplot} shows that the model number count function fails to fit the observed number counts well for magnitudes $I\simeq 15.4$. In order to convert stellar density to a distribution of apparent
magnitude, the relevant quantity is $\rho r^{3}$ \citep{2002MNRAS.330..591B}. Depending on the line-of-sight, this quantity can give asymmetric
magnitude distributions through the bulge. Using the best-fitting analytic
tri-axial density models for the bulge (Rattenbury et al. 2007, in preparation),
this asymmetry is observed and may explain the excess of stars in the
number count histograms, compared to the best-fitting two-component fit
of equation~(\ref{eq:count}). The inability of equation~(\ref{eq:count}) to model completely all features in the observed number counts in some cases leads to an additional uncertainty
in the magnitude location of the fitted Gaussian peak. Computing the
apparent magnitude distribution as $\propto \rho r^{3}$ also produces a
small shift in the peak of the magnitude distribution. This shift is
$\sim +0.04$ mag for $l = 0\deg$, $b = 0\deg$. The proper motion dispersions computed here are
unlikely to be sensitive to these small offsets.

\begin{table*}
\caption{\label{tab:fieldconsts}Values of fitted parameters in equation~(\ref{eq:count}) for all 45 OGLE-II fields used in this analysis. $R$ is the ratio of observed RCG stars to the total number of stars in each field, evaluated over $\pm3\sigma$ around the RCG peak magnitude, $I_{\rm p}$, where $\sigma$ is the fitted Gaussian spread in equation~(\ref{eq:count}). The magnitudes of the model RCG stars are shifted by $\Delta m$ to correspond with the observed mean RCG magnitude in each field. The total number of model stars in each field assigned RCG magnitudes and colours is $n_{\rm rcg}$ and the total number of model stars in each field is $n_{\rm all}$. The corresponding total model weight values for each field are given by $w_{\rm rcg}$ and $w_{\rm all}$ respectively. The large values of $\sigma$ for fields 8-11 might be related to their position at large positive longitudes, and could indicate a structure such as the end of the bar, a ring or spiral arm. An analysis of the bar morphology based on these results is underway (Rattenbury et al. 2007, in preparation).}
\begin{tabular}{lccrrrcrrrrrr}
\hline
Field & $a$ & $b$ & $c$\phantom{xxx} & $I_{\text{p}}$ \phantom{x}& $\sigma$\phantom{x}& $R$&  $\Delta m$ & $n_{\rm rcg}$ & $n_{\rm all}$ & $w_{\rm rcg}$ & $w_{\rm all}$\phantom{x}\\
\hline
 1 & 0.11 & 0.27 & 1735.70 & 14.62 &  0.29 &  0.40 &  0.43 &   585 &  1773 &  277.2 &  842.4 \\ 
 2 & 0.15 & 0.26 & 1876.47 & 14.54 & -0.29 &  0.43 &  0.41 &   621 &  1802 &  298.1 &  853.9 \\ 
 3 & 0.16 & 0.28 & 4692.78 & 14.66 &  0.25 &  0.44 &  0.54 &  1264 &  3626 &  668.5 & 1911.1 \\ 
 4 & 0.17 & 0.28 & 4438.63 & 14.65 &  0.24 &  0.44 &  0.52 &  1298 &  3653 &  670.8 & 1922.2 \\ 
 5 & 0.05 & 0.33 & 4581.59 & 14.70 &  0.28 &  0.33 &  0.55 &  1342 &  4668 &  755.7 & 2685.7 \\ 
 6 & 0.04 & 0.27 &  519.71 & 14.57 &  0.37 &  0.34 &  0.36 &   152 &   583 &   69.5 &  270.8 \\ 
 7 & 0.03 & 0.28 &  457.42 & 14.55 &  0.39 &  0.32 &  0.36 &   143 &   527 &   71.9 &  243.8 \\ 
 8 & 0.04 & 0.27 &  259.65 & 14.37 & -0.51 &  0.22 &  0.35 &    96 &   561 &   41.7 &  236.2 \\ 
 9 & 0.04 & 0.27 &  270.90 & 14.34 &  0.51 &  0.25 & -0.05 &    96 &   497 &   46.1 &  230.9 \\ 
10 & 0.08 & 0.26 &  321.32 & 14.44 &  0.52 &  0.22 &  0.40 &   131 &   654 &   49.1 &  260.1 \\ 
11 & 0.04 & 0.28 &  316.25 & 14.45 &  0.50 &  0.23 &  0.28 &   128 &   695 &   57.5 &  339.4 \\ 
12 & 0.12 & 0.25 &  546.85 & 14.43 &  0.38 &  0.28 &  0.41 &   238 &   908 &  100.7 &  393.1 \\ 
13 & 0.10 & 0.25 &  520.45 & 14.45 &  0.37 &  0.29 &  0.15 &   190 &   863 &   83.9 &  392.4 \\ 
14 & 0.09 & 0.28 & 1309.28 & 14.55 &  0.32 &  0.35 &  0.34 &   458 &  1587 &  216.0 &  767.4 \\ 
15 & 0.05 & 0.29 & 1154.52 & 14.57 &  0.33 &  0.31 &  0.55 &   421 &  1661 &  185.2 &  761.8 \\ 
16 & 0.12 & 0.27 & 1042.72 & 14.50 &  0.35 &  0.33 &  0.50 &   397 &  1383 &  172.8 &  601.1 \\ 
17 & 0.12 & 0.26 & 1069.07 & 14.48 &  0.34 &  0.35 &  0.25 &   406 &  1443 &  212.4 &  753.4 \\ 
18 & 0.17 & 0.26 & 1569.83 & 14.49 &  0.31 &  0.40 &  0.35 &   527 &  1564 &  234.7 &  702.4 \\ 
19 & 0.17 & 0.26 & 1429.23 & 14.51 &  0.32 &  0.40 &  0.44 &   434 &  1365 &  184.4 &  608.5 \\ 
20 & 0.20 & 0.27 & 3012.09 & 14.58 &  0.26 &  0.42 &  0.53 &   939 &  2728 &  480.3 & 1398.3 \\ 
21 & 0.15 & 0.27 & 2793.36 & 14.58 &  0.26 &  0.43 &  0.45 &   900 &  2554 &  443.5 & 1260.0 \\ 
22 & 0.12 & 0.28 & 2574.77 & 14.74 &  0.28 &  0.42 &  0.51 &   830 &  2419 &  382.5 & 1113.3 \\ 
23 & 0.09 & 0.28 & 2147.71 & 14.73 &  0.29 &  0.42 &  0.47 &   767 &  2126 &  384.2 & 1060.6 \\ 
24 & 0.12 & 0.27 & 2130.41 & 14.82 &  0.28 &  0.42 &  0.50 &   595 &  1864 &  269.6 &  905.4 \\ 
25 & 0.07 & 0.28 & 2002.91 & 14.82 &  0.28 &  0.42 &  0.51 &   581 &  1782 &  289.5 &  885.1 \\ 
26 & 0.09 & 0.27 & 1452.89 & 14.83 &  0.31 &  0.38 &  0.55 &   375 &  1325 &  159.7 &  570.5 \\ 
27 & 0.07 & 0.27 & 1319.67 & 14.81 &  0.32 &  0.39 &  0.40 &   387 &  1238 &  172.5 &  578.9 \\ 
28 & 0.04 & 0.28 &  563.00 & 14.79 &  0.31 &  0.31 &  0.62 &   162 &   649 &   72.3 &  293.5 \\ 
29 & 0.05 & 0.27 &  559.86 & 14.78 &  0.31 &  0.32 &  0.44 &   156 &   607 &   70.7 &  267.5 \\ 
30 & 0.18 & 0.27 & 2533.75 & 14.57 &  0.27 &  0.42 &  0.41 &   754 &  2195 &  362.4 & 1026.7 \\ 
31 & 0.17 & 0.27 & 2354.64 & 14.53 &  0.28 &  0.43 &  0.32 &   763 &  2229 &  361.9 & 1122.1 \\ 
32 & 0.17 & 0.26 & 2062.96 & 14.53 &  0.28 &  0.42 &  0.41 &   638 &  1962 &  291.8 &  938.5 \\ 
33 & 0.13 & 0.27 & 1614.83 & 14.56 &  0.31 &  0.41 &  0.34 &   559 &  1586 &  265.5 &  760.7 \\ 
34 & 0.18 & 0.27 & 3210.56 & 14.60 &  0.27 &  0.43 &  0.42 &   990 &  2936 &  503.0 & 1473.9 \\ 
35 & 0.16 & 0.26 & 1963.53 & 14.53 &  0.29 &  0.41 &  0.45 &   663 &  1925 &  307.7 &  913.7 \\ 
36 & 0.16 & 0.26 & 1773.62 & 14.51 &  0.30 &  0.41 &  0.47 &   574 &  1902 &  301.1 &  943.5 \\ 
37 & 0.18 & 0.28 & 4901.22 & 14.64 &  0.25 &  0.42 &  0.43 &  1439 &  4077 &  794.9 & 2218.5 \\ 
38 & 0.12 & 0.27 & 2091.19 & 14.64 &  0.28 &  0.43 &  0.46 &   662 &  1945 &  319.2 &  948.1 \\ 
39 & 0.18 & 0.28 & 3919.30 & 14.69 &  0.26 &  0.44 &  0.65 &  1217 &  3456 &  631.8 & 1804.2 \\ 
40 & 0.09 & 0.28 & 2181.18 & 14.87 &  0.29 &  0.41 &  0.62 &   668 &  1936 &  315.1 &  933.3 \\ 
41 & 0.10 & 0.28 & 2180.49 & 14.87 &  0.28 &  0.42 &  0.55 &   626 &  1905 &  318.2 &  965.4 \\ 
42 & 0.13 & 0.26 & 1215.38 & 14.52 &  0.35 &  0.37 &  0.40 &   425 &  1389 &  190.2 &  637.7 \\ 
43 & 0.10 & 0.28 & 2659.91 & 14.84 &  0.27 &  0.41 &  0.79 &   777 &  2290 &  345.8 & 1074.6 \\ 
45 & 0.11 & 0.27 & 1541.36 & 14.59 &  0.31 &  0.40 &  0.38 &   485 &  1568 &  228.3 &  767.7 \\ 
46 & 0.09 & 0.27 & 1428.63 & 14.60 &  0.30 &  0.41 &  0.38 &   454 &  1400 &  221.6 &  669.5 \\ 

\hline
\end{tabular}

\end{table*}

Each star in the galactic model is assigned a RCG magnitude  with probability $R_{k}$ for each field. The
apparent magnitude is computed using the model distance. Stars which
are not assigned a RCG magnitude are assigned a magnitude using the
power-law component of equation~(\ref{eq:count}), defined over the same
limits used to compute $R_{k}$. Here we implicitly assume that the RCG
stars trace the overall Galactic disk and bulge populations.

The RCG luminosity function is approximated by a Gaussian distribution with  mean magnitude $-0.26$ and $\sigma = 0.2$. These assumptions are  mostly consistent with observations \citep{1997ApJ...477..163S} and the fitted distribution from \citet{2000ApJ...531L..25U}, but there may be small offsets between local and bulge red clump giants. It was noted in
\citet{2004MNRAS.349..193S} that there is some as-yet unexplained
offset (0.3 mag) in the
extinction-corrected mean RCG magnitudes in the OGLE fields. A possible explanation for this offset is that the RCG population effects are large: so that the absolute magnitude of RCG stars is significantly different for RCGs in the bulge compared to local RCGs, as claimed by \citet{2003MNRAS.343..539P} and \citet{2003ApJ...588..801S}. A different value of the distance to the Galactic centre to that assumed here (8 \kpc) would in part account for the discrepancy, however would not remove it completely. Using a value of 7.6 \kpc \citep{2005ApJ...628..246E, 2006ApJ...647.1093N} as the distance to the Galactic centre would change the zero-point by 0.12 mag, resulting in an offset value of 0.18 mag. It is also possible that reddening toward the Galactic centre is more complicated than assumed in \citet{2004MNRAS.349..193S}. In order
to compare the model proper motion results with the observed data, it
was necessary to shift the mean model RCG magnitudes to correspond
with that observed in each of the OGLE fields. The model RCG
magnitudes were fitted with a Gaussian curve. The mean of the model
RCG magnitudes was then shifted by a value $\Delta m$, see Table~\ref{tab:fieldconsts}, to correspond with the
observed mean RCG magnitude in each of the OGLE fields.  Notice that
we concentrate on second-order moments (proper motion dispersions) of the proper
motion, so a small shift in the zero-point has little effect on our
results.

Every model particle has an associated weight, $w_{i}$. The particle
weight can take values $0 < w_{i} \lesssim 20$.  In order to account
for this weighting, $\lceil w_{i} \rceil$ stars are generated for each
particle with the same kinematics but magnitudes determined as
above. $\lceil w_{i} \rceil$ is the nearest integer toward $+\infty$.
Each model star is then assigned a weight, $\gamma_{i} = w_{i}/\lceil
w_{i} \rceil$. Notice this procedure allows us to increase the
effective number of particles to better sample the luminosity
function.  The total number of stars and the number of stars assigned RCG magnitudes in each field are listed in Table~\ref{tab:fieldconsts} as $n_{\rm all}$ and $n_{\rm rcg}$ respectively. 81806 stars from the model were used to compare model kinematics to observed values.

\subsection{Model kinematics}
\label{sec:modkine}

Stars with apparent magnitudes within the limits $m_{\rm min} = 13.7$ and $m_{\rm max} = 15.5$, were selected from the model data. This magnitude range corresponds to the selection criteria imposed on the observed data sample, see section~\ref{sec:observed}. Model stars with total proper motions greater than 10 \masyr (corresponding to $>380\kms$ at a distance of the Galactic centre) were excluded on the basis that such stars would be similarly excluded from any observed sample. The fraction of weight removed and number of stars removed in this way only amounted to a few per cent of the total weight and number of stars in each field. Bulge model stars were selected by requiring a distance $d>6$ \kpc.

The mean proper motion and proper motion dispersions in the latitude and longitude directions were computed along with their errors for all model stars in each field which obey the above selection criteria. The weights on model stars, $\gamma_{i}$, were used to compute these values.

We then tested whether the finite and discrete nature of the model data gives rise to uncertainties in the measured proper motion dispersion values. We measured the intrinsic noise in the model by comparing the proper motion dispersions computed for four equivalent lines-of-sight through the model for each field. The spread of the proper motion dispersions for each field was then used as the estimate of the intrinsic noise in the model. The mean (median) value of these errors in the longitude and latitude directions are  0.08 (0.06) and 0.12 (0.097) \masyr respectively.

The statistical error for the proper motion dispersions in the longitude and latitude directions for each field were combined in quadrature with the  error arising from the finite discrete nature of the model data to give the total error on the proper motion dispersions computed from the model.

\section{Comparison between theoretical model and observed data}
\label{sec:results}
\begin{figure*}
\setlength{\voff}{-5pt}
\psfrag{xlabel1}{\hspace{-20pt}\raisebox{\voff}{\normalsize Model $\sigma_{\rm l}$ (\masyr)}}
\psfrag{ylabel1}{\hspace{-40pt}\normalsize{Observed $\sigma_{\rm l}$ (\masyr)}}
\psfrag{xlabel2}{\hspace{-20pt}\raisebox{\voff}{\normalsize Model $\sigma_{\rm b}$ (\masyr) }}
\psfrag{ylabel2}{\hspace{-40pt}\normalsize{Observed $\sigma_{\rm b}$ (\masyr)}}
\psfrag{XXXlabel1}{\scriptsize BW fields}
\psfrag{XXXlabel2}{$ \scriptstyle |l| > 5\deg$}
\psfrag{XXXlabel3}{$\scriptstyle |l| < 5\deg$}
\psfrag{YYYlabel1}{\scriptsize BW fields}
\psfrag{YYYlabel2}{$ \scriptstyle |l| > 5\deg$}
\psfrag{YYYlabel3}{$\scriptstyle |l| < 5\deg$}

\centering\includegraphics[width=0.49\hsize]{\FigDir{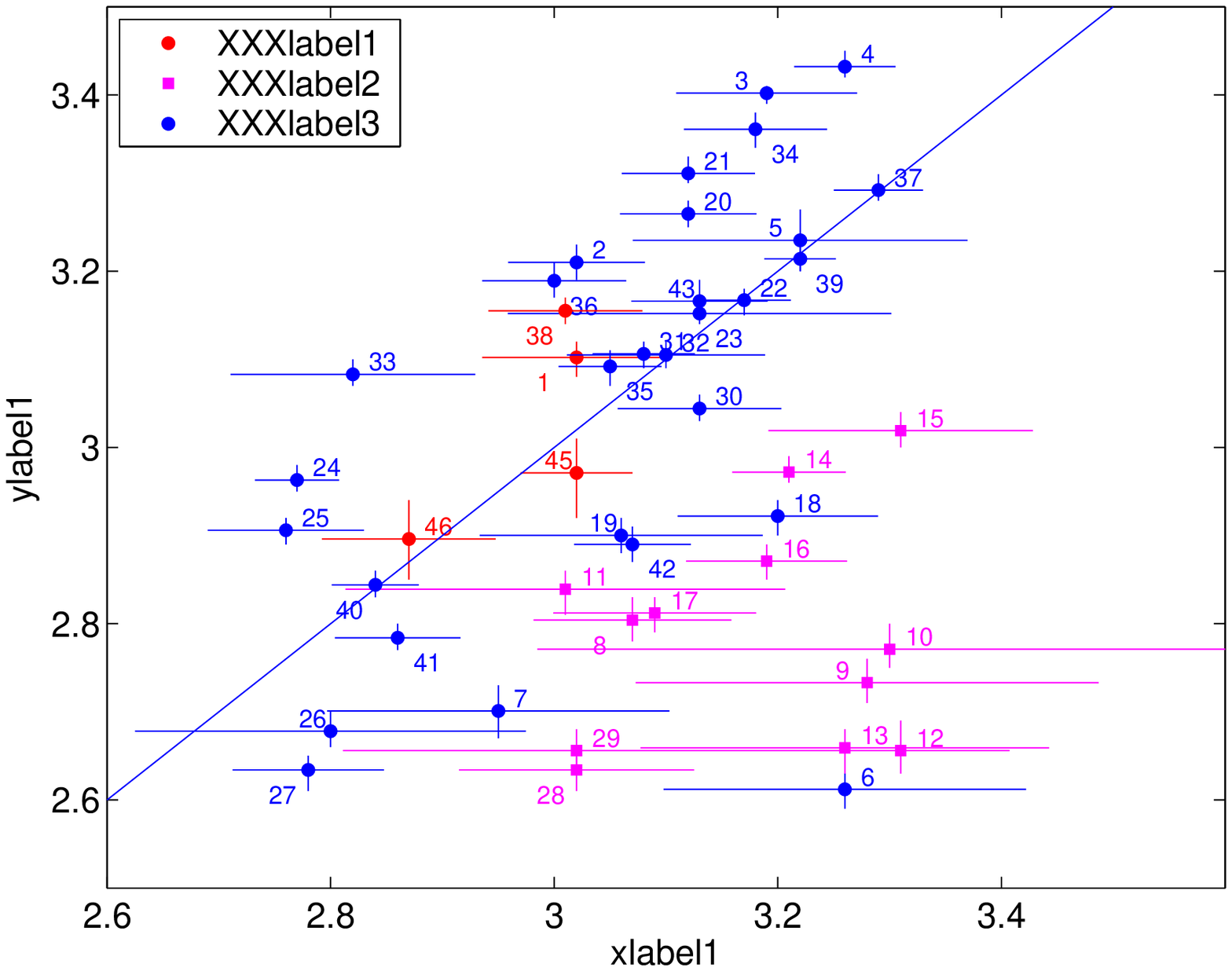}}
\centering\includegraphics[width=0.49\hsize]{\FigDir{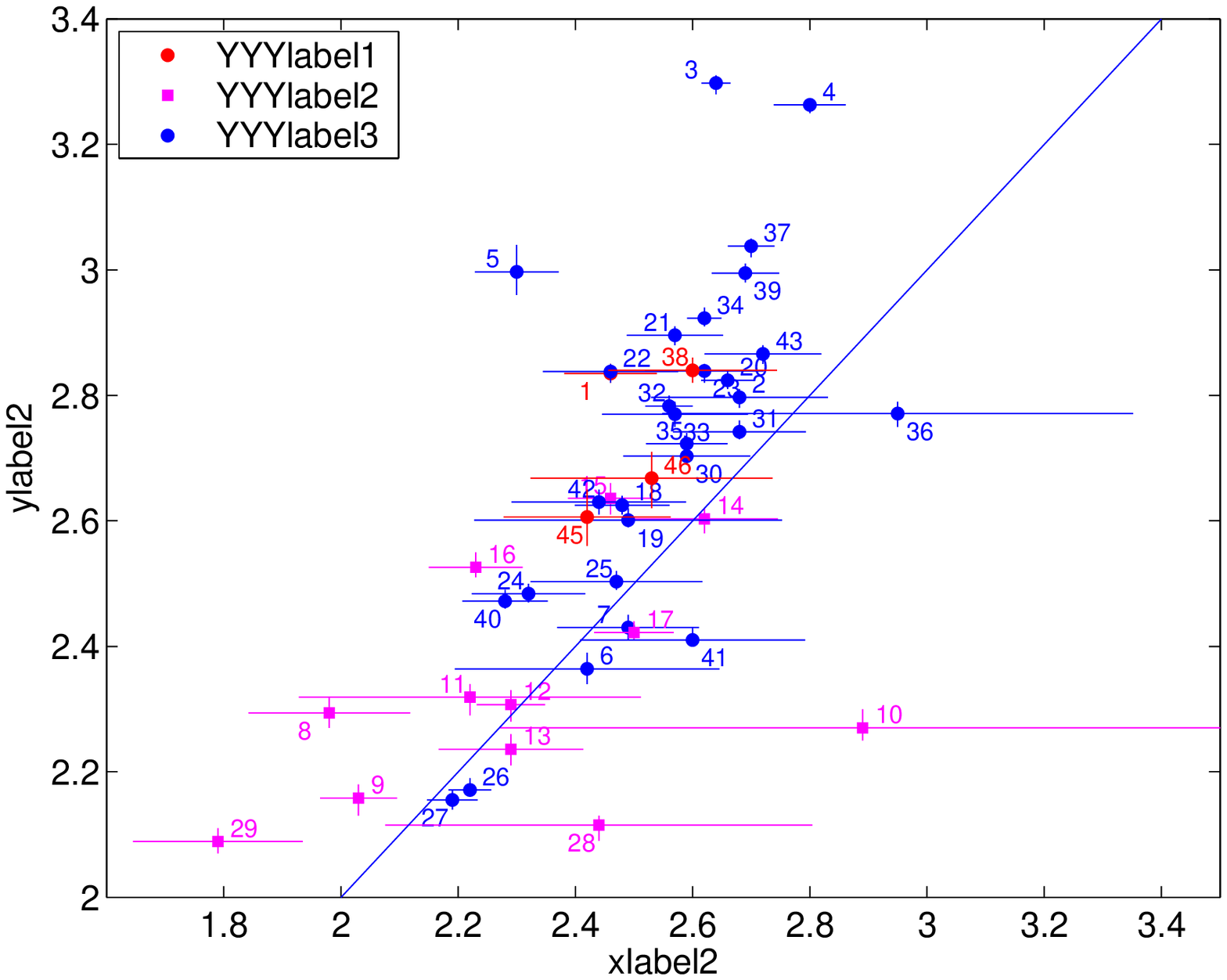}}
\caption{Comparison between observed and predicted proper motion dispersions for
stars in the OGLE-II proper motion catalogue of
\citet{2004MNRAS.348.1439S}. Left: Proper motion dispersions in the galactic
longitude direction, $\sigma_{\rm l}$. The OGLE-II field number is
indicated adjacent to each point, see also
Fig.~\ref{fig:fields}. Fields with galactic longitude $|l| >
5\deg$ are shown in magenta; fields within Baade's window are
shown in red; all other fields in blue. Right: Proper motion dispersions in the
galactic latitude direction, $\sigma_{\rm b}$, shown with the same colour
scheme.}
\label{fig:dispersions}

\end{figure*}

The observed and  predicted proper motion dispersions for each of the OGLE-II fields  are shown in Table~\ref{tab:theresults}.  Fig.~\ref{fig:dispersions} shows the observed  proper motion dispersions for each of the analysed OGLE-II fields plotted against the predicted model proper motion dispersions.

Fig.~\ref{fig:dispersions} shows that the model predictions are in general agreement with observed proper motion dispersions for the OGLE-II fields. The model has been used previously  to predict the proper motion dispersions of 427 stars\footnote{There are two repeated entries in Table~2 of \citet{1992AJ....103..297S}.} entries observed by \citet{1992AJ....103..297S} in a single 6\arcmin $\times$ 6\arcmin \ field toward the bulge \citep{2004ApJ...601L.155B}. The model value of $\sigma_{\rm l}$ in this previous analysis was in agreement with the observed value, yet the model and observed values of $\sigma_{\rm b}$ were significantly different. The 6\arcmin $\times$ 6\arcmin \ field used by \citet{1992AJ....103..297S} falls within the OGLE-II field number 45. The model prediction of $\sigma_{\rm l}$ for stars in OGLE field 45 is completely consistent with the measured value. The model prediction of $\sigma_{\rm b}$ shows a similar discrepancy to the previous analysis of \citet{2004ApJ...601L.155B}.

\begin{figure*}
\setlength{\voff}{-5pt}
\psfrag{xlabel3}{\hspace{-20pt}\raisebox{\voff}{\normalsize{Model $\sigma_{\rm l} / \sigma_{\rm b}$}  }}
\psfrag{ylabel3}{\hspace{-15pt}\normalsize{Observed $\sigma_{\rm l} / \sigma_{\rm b}$}}
\psfrag{xlabel4}{\hspace{-10pt}\raisebox{\voff}{\normalsize{Model $C_{\rm lb}$}}}
\psfrag{ylabel4}{\hspace{-10pt}\raisebox{-5pt}{\normalsize{Observed $C_{\rm lb}$}}}
\psfrag{YYYlabel1}{\scriptsize BW fields}
\psfrag{YYYlabel2}{$ \scriptstyle |l| > 5\deg$}
\psfrag{YYYlabel3}{$\scriptstyle |l| < 5\deg$}
\psfrag{ZZZlabel1}{\scriptsize BW fields}
\psfrag{ZZZlabel2}{$ \scriptstyle |l| > 5\deg$}
\psfrag{ZZZlabel3}{$\scriptstyle |l| < 5\deg$}
\centering\includegraphics[width=0.49\hsize]{\FigDir{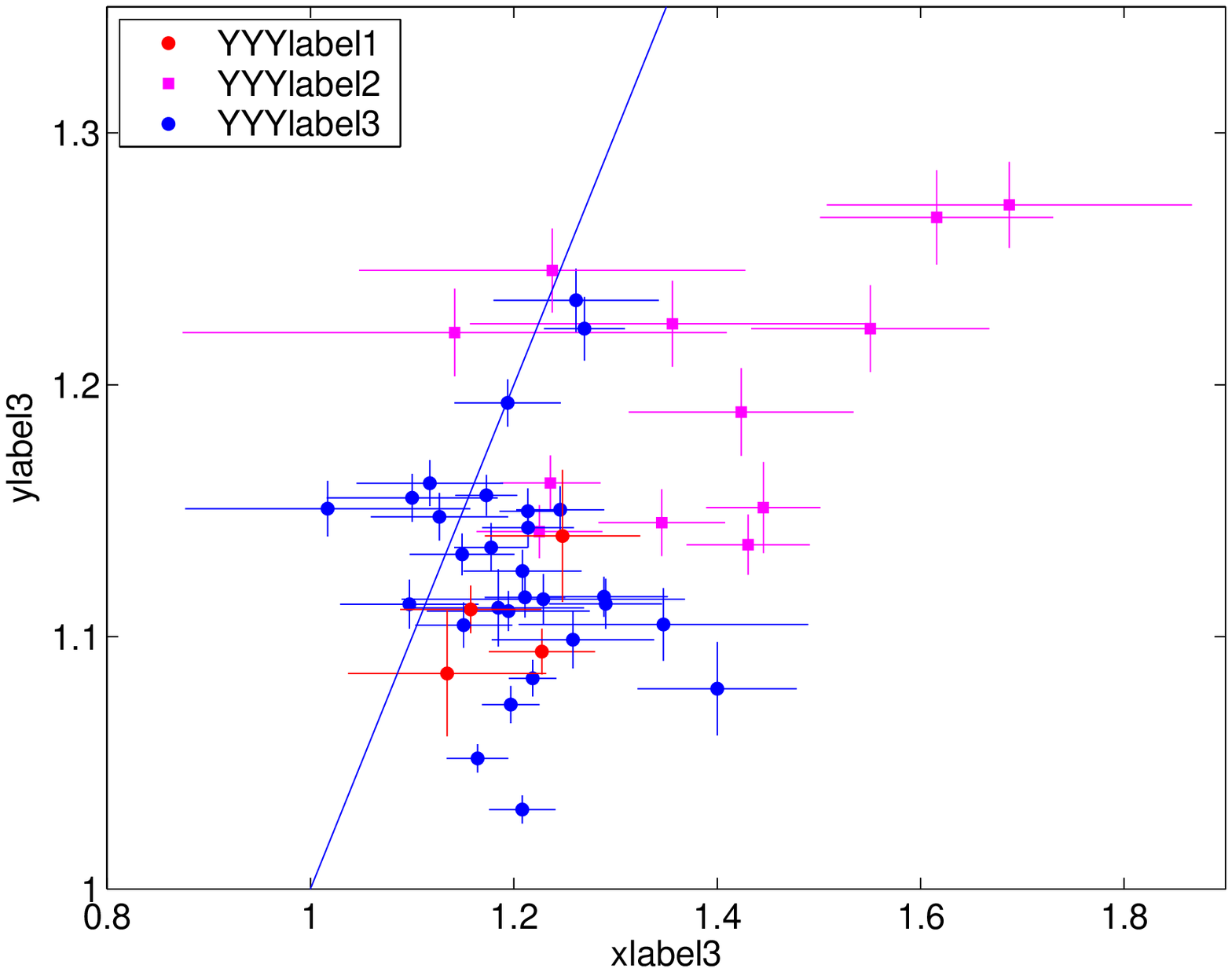}}
\centering\includegraphics[width=0.5\hsize]{\FigDir{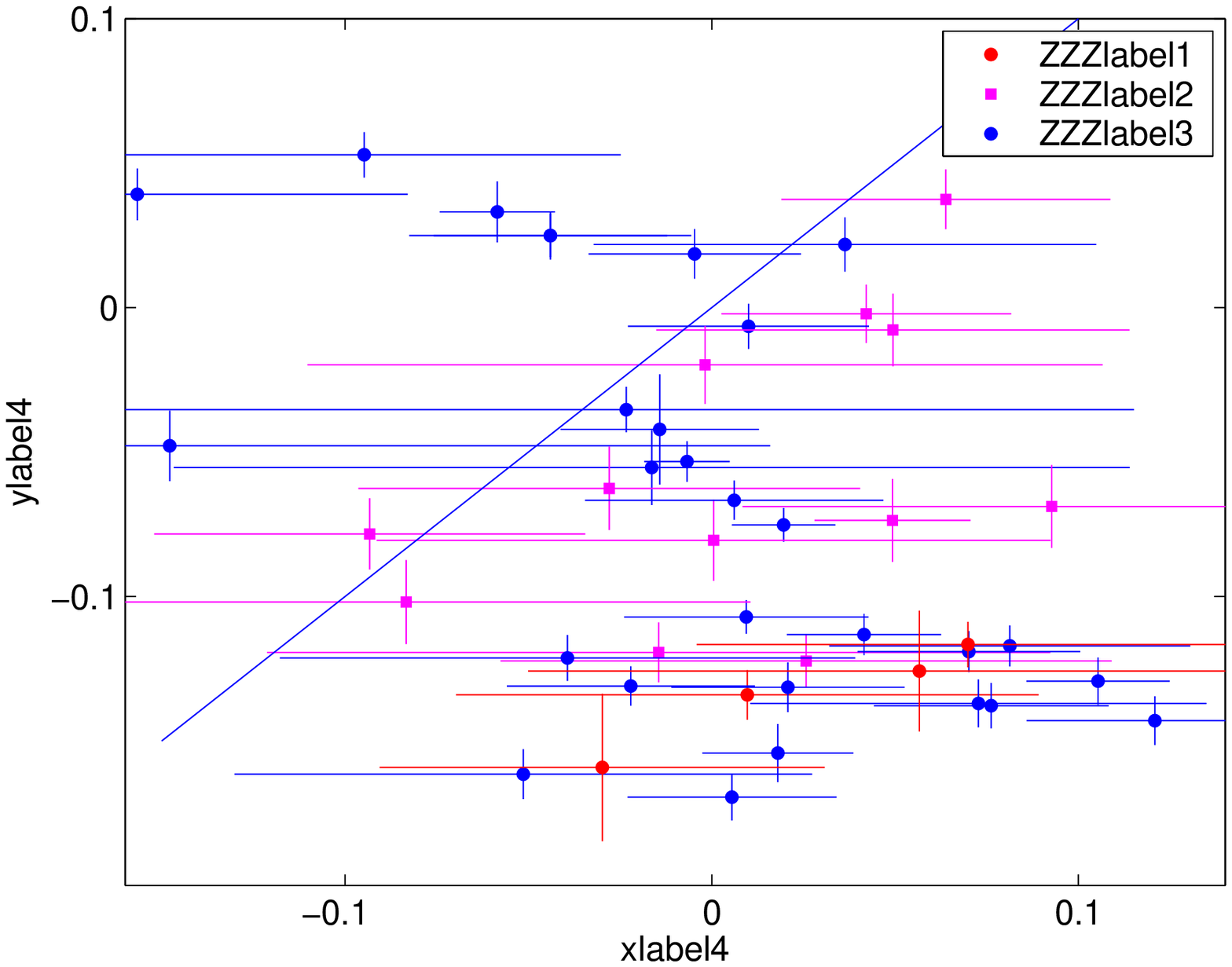}}
\caption{Left: Ratio of proper motion dispersions $R = \sigma_{\rm l} / \sigma_{\rm b}$ for the observed OGLE-II proper motion data and model predictions. The model generally predicts more anisotropic motion, i.e. $R>1$ than is observed in the data. Right: The cross-correlation term $C_{\rm lb} = \sigma_{\rm lb} / \sigma_{\rm l}\sigma_{\rm b}$. }
\label{fig:ratios}

\end{figure*}

Fig.~\ref{fig:ratios} shows the ratio $R=\sigma_{\rm l} / \sigma_{\rm b}$ and cross-correlation term $C_{\rm lb} = \sigma_{\rm lb} / (\sigma_{\rm l}\sigma_{\rm b})$ computed using the model and observed data. Typically the model predicts more anisotropic motion with $R>1$ than what is observed.  

The model predictions for stellar kinematics in the latitude direction may be problematic. This is not surprising as the model is not well constrained toward the plane due to a lack of observational data because of the heavy dust extinction. The problem is currently under investigation. Similarly, the model predictions for $\sigma_{\rm l}$ degrade as $l$ increases. This is because the model performance has been optimised for regions close to the Galactic centre.

The significant difference between the observed proper motion dispersions of adjacent fields (e.g. fields 1 and 45) might hint at some fine-scale population effect, whereby a group of stars surviving the selection criteria have a significant and discrepant kinematic signature. Higher-accuracy observations using the HST support this evidence of such population effects \citep{2006MNRAS.370..435K}.

No attempt has been made to account for the blending of flux inherent
in the OGLE-II crowded-field photometry. It is certain that a fraction
of stars in each OGLE-II field suffers from some degree of blending
\citep{2006MNRAS.370..435K}. To investigate this effect, we checked one
field covering the lens MACHO-95-BLG-37 ($l=2.54\deg,
b=3.33\deg$, \citealt{2005ApJ...631..906T}) from the HST proper
motion survey of \citet{2006MNRAS.370..435K}, which falls inside
OGLE-II field number 2. HST images suffer much less blending, but the
field of view is small, and so it has only a dozen or so clump
giants. We derive a proper motion of $\sigma_{\rm l}=3.13\pm 0.57 $ \masyr,
and $\sigma_{\rm b}=2.17\pm 0.40 $ \masyr. These values agree with our
kinematics in field 2 within $0.2\sigma$ for $\sigma_{\rm l}$ and $1.6\sigma$ 
for $\sigma_{\rm b}$. The errors in
our proper motion dispersions are very small ($\sim \kms$ at a
distance of the Galactic centre), but it is likely that we
underestimate the error bars on the observed data due to systematic
effects such as blending.

\begin{table*}
\caption{\label{tab:theresults}Proper motion dispersions in the longitude and latitude directions, $\sigma_{\rm l}$, $\sigma_{\rm b}$ , and cross-correlation term $C_{\rm lb}$ for bulge stars in 45 OGLE-II fields. High precision proper motion data for bulge stars were
extracted from the OGLE-II proper motion catalogue
\citep{2004MNRAS.348.1439S}.  $N$ is the number of stars selected from each field. Field 44 was not used due to the low
number of RCGs in this field.}
  \begin{tabular}{crrccccrrr}
\hline 
\multicolumn{3}{c}{} &  \multicolumn{4}{c}{PM Dispersions (\masyr)} &  \multicolumn{2}{c}{$C_{\rm lb}$}\\

Field & \multicolumn{2}{c}{Field centre}  & \multicolumn{2}{c}{Longitude $\sigma_{\rm l}$} & \multicolumn{2}{c}{Latitude $\sigma_{\rm b}$} & \multicolumn{2}{c}{} & \\

& $l\,(\deg)$ & $b\,(\deg)$  & {Model} & {Observed} & {Model} & {Observed} & {Model}\phantom{Xx}& {Observed}\phantom{X} & $N$\phantom{X}\\
\hline
1 & 1.08 & -3.62 & 3.02  \ppm  0.08 & 3.10  \ppm  0.02 & 2.46  \ppm  0.08 & 2.83  \ppm  0.02 & 0.01  \ppm  0.08 & -0.13  \ppm  0.01 & 15434 \\ 
2 & 2.23 & -3.46 & 3.02  \ppm  0.06 & 3.21  \ppm  0.02 & 2.68  \ppm  0.15 & 2.80  \ppm  0.02 & 0.12  \ppm  0.03 & -0.14  \ppm  0.01 & 16770 \\ 
3 & 0.11 & -1.93 & 3.19  \ppm  0.08 & 3.40  \ppm  0.01 & 2.64  \ppm  0.02 & 3.30  \ppm  0.02 & 0.02  \ppm  0.01 & -0.08  \ppm  0.01 & 26763 \\ 
4 & 0.43 & -2.01 & 3.26  \ppm  0.05 & 3.43  \ppm  0.02 & 2.80  \ppm  0.06 & 3.26  \ppm  0.01 & 0.01  \ppm  0.03 & -0.11  \ppm  0.01 & 26382 \\ 
5 & -0.23 & -1.33 & 3.22  \ppm  0.15 & 3.23  \ppm  0.03 & 2.30  \ppm  0.07 & 3.00  \ppm  0.04 & -0.01  \ppm  0.03 & -0.04  \ppm  0.02 & 3145 \\ 
6 & -0.25 & -5.70 & 3.26  \ppm  0.16 & 2.61  \ppm  0.02 & 2.42  \ppm  0.23 & 2.36  \ppm  0.03 & -0.02  \ppm  0.13 & -0.06  \ppm  0.01 & 7027 \\ 
7 & -0.14 & -5.91 & 2.95  \ppm  0.15 & 2.70  \ppm  0.03 & 2.49  \ppm  0.12 & 2.43  \ppm  0.02 & -0.15  \ppm  0.16 & -0.05  \ppm  0.01 & 6236 \\ 
8 & 10.48 & -3.78 & 3.07  \ppm  0.09 & 2.80  \ppm  0.03 & 1.98  \ppm  0.14 & 2.29  \ppm  0.02 & 0.00  \ppm  0.09 & -0.08  \ppm  0.01 & 5136 \\ 
9 & 10.59 & -3.98 & 3.28  \ppm  0.21 & 2.73  \ppm  0.02 & 2.03  \ppm  0.07 & 2.16  \ppm  0.03 & -0.03  \ppm  0.07 & -0.06  \ppm  0.01 & 5114 \\ 
10 & 9.64 & -3.44 & 3.30  \ppm  0.32 & 2.77  \ppm  0.02 & 2.89  \ppm  0.62 & 2.27  \ppm  0.02 & 0.09  \ppm  0.08 & -0.07  \ppm  0.01 & 5568 \\ 
11 & 9.74 & -3.64 & 3.01  \ppm  0.20 & 2.84  \ppm  0.02 & 2.22  \ppm  0.29 & 2.32  \ppm  0.02 & -0.08  \ppm  0.09 & -0.10  \ppm  0.01 & 5369 \\ 
12 & 7.80 & -3.37 & 3.31  \ppm  0.10 & 2.66  \ppm  0.03 & 2.29  \ppm  0.06 & 2.31  \ppm  0.03 & -0.09  \ppm  0.06 & -0.08  \ppm  0.01 & 6035 \\ 
13 & 7.91 & -3.58 & 3.26  \ppm  0.18 & 2.66  \ppm  0.03 & 2.29  \ppm  0.12 & 2.24  \ppm  0.02 & 0.05  \ppm  0.02 & -0.07  \ppm  0.01 & 5601 \\ 
14 & 5.23 & 2.81 & 3.21  \ppm  0.05 & 2.97  \ppm  0.02 & 2.62  \ppm  0.13 & 2.60  \ppm  0.02 & 0.06  \ppm  0.04 & 0.04  \ppm  0.01 & 10427 \\ 
15 & 5.38 & 2.63 & 3.31  \ppm  0.12 & 3.02  \ppm  0.02 & 2.46  \ppm  0.07 & 2.64  \ppm  0.03 & 0.04  \ppm  0.04 & -0.00  \ppm  0.01 & 8989 \\ 
16 & 5.10 & -3.29 & 3.19  \ppm  0.07 & 2.87  \ppm  0.02 & 2.23  \ppm  0.08 & 2.53  \ppm  0.02 & 0.03  \ppm  0.08 & -0.12  \ppm  0.01 & 9799 \\ 
17 & 5.28 & -3.45 & 3.09  \ppm  0.09 & 2.81  \ppm  0.02 & 2.50  \ppm  0.07 & 2.42  \ppm  0.01 & -0.01  \ppm  0.11 & -0.12  \ppm  0.01 & 10268 \\ 
18 & 3.97 & -3.14 & 3.20  \ppm  0.09 & 2.92  \ppm  0.02 & 2.48  \ppm  0.08 & 2.62  \ppm  0.02 & 0.02  \ppm  0.03 & -0.13  \ppm  0.01 & 14019 \\ 
19 & 4.08 & -3.35 & 3.06  \ppm  0.13 & 2.90  \ppm  0.02 & 2.49  \ppm  0.26 & 2.60  \ppm  0.02 & 0.01  \ppm  0.03 & -0.17  \ppm  0.01 & 13256 \\ 
20 & 1.68 & -2.47 & 3.12  \ppm  0.06 & 3.27  \ppm  0.01 & 2.66  \ppm  0.05 & 2.82  \ppm  0.01 & 0.07  \ppm  0.03 & -0.12  \ppm  0.01 & 17678 \\ 
21 & 1.80 & -2.66 & 3.12  \ppm  0.06 & 3.31  \ppm  0.02 & 2.57  \ppm  0.08 & 2.90  \ppm  0.02 & -0.02  \ppm  0.03 & -0.13  \ppm  0.01 & 17577 \\ 
22 & -0.26 & -2.95 & 3.17  \ppm  0.04 & 3.17  \ppm  0.02 & 2.46  \ppm  0.12 & 2.84  \ppm  0.02 & 0.01  \ppm  0.03 & -0.01  \ppm  0.01 & 19787 \\ 
23 & -0.50 & -3.36 & 3.13  \ppm  0.17 & 3.15  \ppm  0.01 & 2.62  \ppm  0.10 & 2.84  \ppm  0.02 & -0.02  \ppm  0.14 & -0.04  \ppm  0.01 & 17996 \\ 
24 & -2.44 & -3.36 & 2.77  \ppm  0.04 & 2.96  \ppm  0.01 & 2.32  \ppm  0.10 & 2.48  \ppm  0.01 & -0.04  \ppm  0.04 & 0.02  \ppm  0.01 & 16397 \\ 
25 & -2.32 & -3.56 & 2.76  \ppm  0.07 & 2.91  \ppm  0.01 & 2.47  \ppm  0.15 & 2.50  \ppm  0.01 & -0.04  \ppm  0.03 & 0.02  \ppm  0.01 & 16386 \\ 
26 & -4.90 & -3.37 & 2.80  \ppm  0.17 & 2.68  \ppm  0.02 & 2.22  \ppm  0.04 & 2.17  \ppm  0.01 & -0.00  \ppm  0.03 & 0.02  \ppm  0.01 & 13099 \\ 
27 & -4.92 & -3.65 & 2.78  \ppm  0.07 & 2.63  \ppm  0.02 & 2.19  \ppm  0.04 & 2.15  \ppm  0.01 & -0.06  \ppm  0.02 & 0.03  \ppm  0.01 & 12728 \\ 
28 & -6.76 & -4.42 & 3.02  \ppm  0.11 & 2.63  \ppm  0.03 & 2.44  \ppm  0.36 & 2.12  \ppm  0.02 & 0.05  \ppm  0.06 & -0.01  \ppm  0.01 & 8367 \\ 
29 & -6.64 & -4.62 & 3.02  \ppm  0.21 & 2.66  \ppm  0.03 & 1.79  \ppm  0.14 & 2.09  \ppm  0.02 & -0.00  \ppm  0.11 & -0.02  \ppm  0.01 & 8108 \\ 
30 & 1.94 & -2.84 & 3.13  \ppm  0.07 & 3.04  \ppm  0.02 & 2.59  \ppm  0.11 & 2.70  \ppm  0.02 & -0.04  \ppm  0.08 & -0.12  \ppm  0.01 & 17774 \\ 
31 & 2.23 & -2.94 & 3.08  \ppm  0.05 & 3.11  \ppm  0.02 & 2.68  \ppm  0.11 & 2.74  \ppm  0.01 & 0.08  \ppm  0.05 & -0.12  \ppm  0.01 & 17273 \\ 
32 & 2.34 & -3.14 & 3.10  \ppm  0.09 & 3.10  \ppm  0.02 & 2.56  \ppm  0.04 & 2.78  \ppm  0.01 & 0.11  \ppm  0.02 & -0.13  \ppm  0.01 & 15966 \\ 
33 & 2.35 & -3.66 & 2.82  \ppm  0.11 & 3.08  \ppm  0.02 & 2.57  \ppm  0.12 & 2.77  \ppm  0.02 & 0.07  \ppm  0.06 & -0.14  \ppm  0.01 & 15450 \\ 
34 & 1.35 & -2.40 & 3.18  \ppm  0.06 & 3.36  \ppm  0.02 & 2.62  \ppm  0.03 & 2.92  \ppm  0.01 & 0.04  \ppm  0.02 & -0.11  \ppm  0.01 & 16889 \\ 
35 & 3.05 & -3.00 & 3.05  \ppm  0.05 & 3.09  \ppm  0.02 & 2.59  \ppm  0.07 & 2.72  \ppm  0.02 & 0.08  \ppm  0.03 & -0.14  \ppm  0.01 & 15973 \\ 
36 & 3.16 & -3.20 & 3.00  \ppm  0.06 & 3.19  \ppm  0.02 & 2.95  \ppm  0.40 & 2.77  \ppm  0.02 & -0.05  \ppm  0.08 & -0.16  \ppm  0.01 & 14955 \\ 
37 & 0.00 & -1.74 & 3.29  \ppm  0.04 & 3.29  \ppm  0.02 & 2.70  \ppm  0.04 & 3.04  \ppm  0.01 & -0.01  \ppm  0.01 & -0.05  \ppm  0.01 & 20233 \\ 
38 & 0.97 & -3.42 & 3.01  \ppm  0.07 & 3.15  \ppm  0.01 & 2.60  \ppm  0.14 & 2.84  \ppm  0.02 & 0.07  \ppm  0.07 & -0.12  \ppm  0.01 & 15542 \\ 
39 & 0.53 & -2.21 & 3.22  \ppm  0.03 & 3.21  \ppm  0.01 & 2.69  \ppm  0.06 & 3.00  \ppm  0.01 & 0.01  \ppm  0.04 & -0.07  \ppm  0.01 & 24820 \\ 
40 & -2.99 & -3.14 & 2.84  \ppm  0.04 & 2.84  \ppm  0.01 & 2.28  \ppm  0.07 & 2.47  \ppm  0.02 & -0.09  \ppm  0.07 & 0.05  \ppm  0.01 & 13581 \\ 
41 & -2.78 & -3.27 & 2.86  \ppm  0.06 & 2.78  \ppm  0.01 & 2.60  \ppm  0.19 & 2.41  \ppm  0.02 & -0.16  \ppm  0.07 & 0.04  \ppm  0.01 & 14070 \\ 
42 & 4.48 & -3.38 & 3.07  \ppm  0.05 & 2.89  \ppm  0.02 & 2.44  \ppm  0.15 & 2.63  \ppm  0.02 & 0.02  \ppm  0.02 & -0.15  \ppm  0.01 & 10099 \\ 
43 & 0.37 & 2.95 & 3.13  \ppm  0.06 & 3.17  \ppm  0.02 & 2.72  \ppm  0.10 & 2.87  \ppm  0.01 & 0.04  \ppm  0.07 & 0.02  \ppm  0.01 & 11467 \\ 
45 & 0.98 & -3.94 & 3.02  \ppm  0.05 & 2.97  \ppm  0.04 & 2.42  \ppm  0.14 & 2.61  \ppm  0.04 & 0.06  \ppm  0.11 & -0.13  \ppm  0.02 & 2380 \\ 
46 & 1.09 & -4.14 & 2.87  \ppm  0.08 & 2.90  \ppm  0.04 & 2.53  \ppm  0.21 & 2.67  \ppm  0.04 & -0.03  \ppm  0.06 & -0.16  \ppm  0.03 & 1803 \\ 

\hline
\end{tabular}

\end{table*}

\subsection{Understanding the differences}
\label{sec:diff}
\begin{figure}
\psfrag{ylabel1}{\normalsize \hspace{-25pt}$\Delta_{\rm l,obs}$ (\masyr)}
\psfrag{ylabel2}{\normalsize \hspace{-25pt} $\Delta_{\rm b,obs}$ (\masyr)}
\psfrag{xlabel}{\normalsize \raisebox{-2pt}{\hspace{-12pt} Separation $(^{\circ})$}}
\centering\includegraphics[width=1.0\hsize]{\FigDir{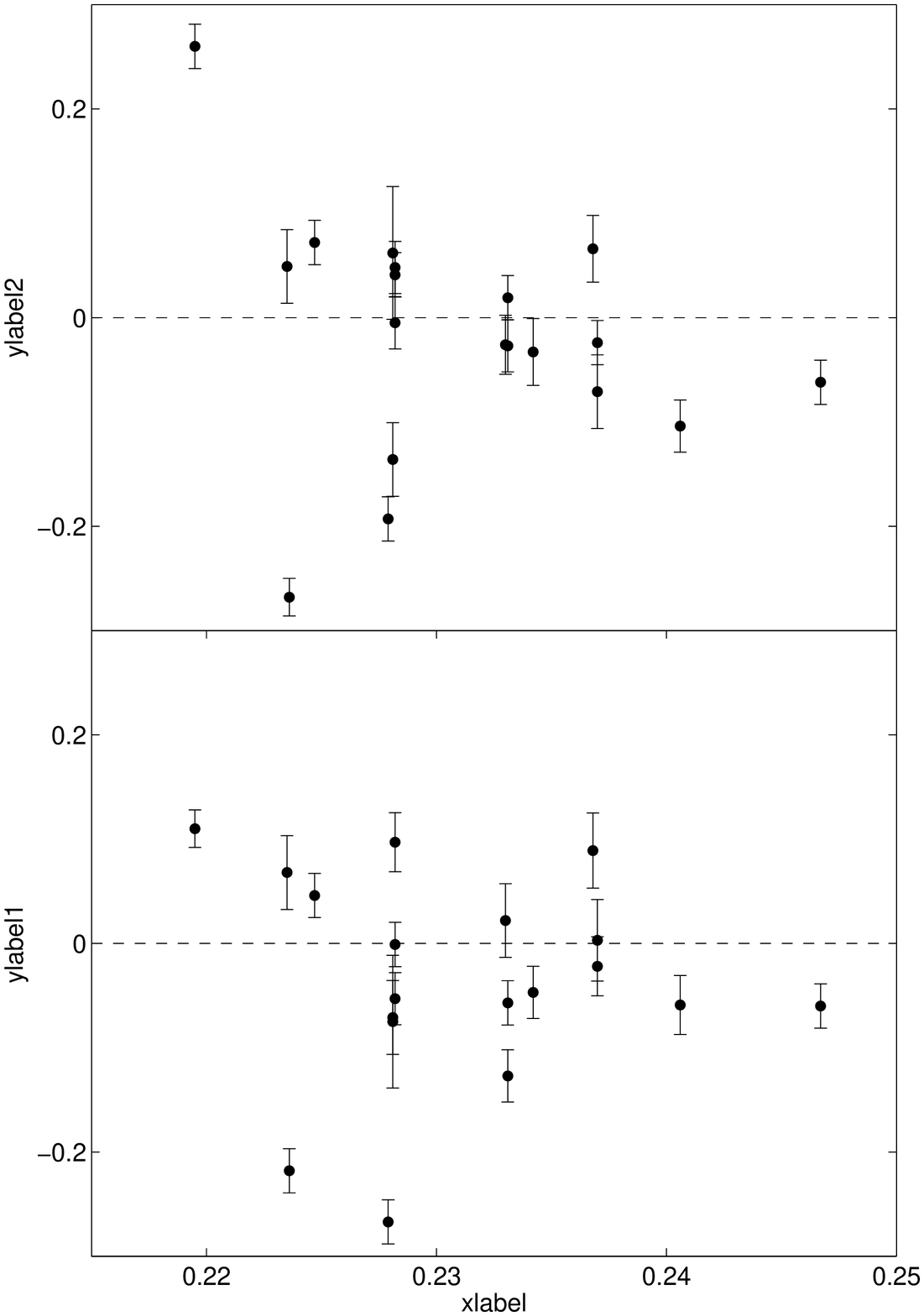}}
\caption{\label{fig:ObsDispersionResiduals}Difference between observed proper motion dispersions  for pairs of fields with separations less than 0.25 degrees (corresponding to $\simeq 40$ pc at the Galactic centre).}
\end{figure}

We now seek to understand the cause of the differences between the
model and the Milky Way, at least at a qualitative level.  We first consider
the possibility that the difference can be explained by some
systematic effect.  We compute the differences between observed proper motion dispersions of nearest fields
for fields with separations less than $0.25$ degrees.  No pair of
fields is used twice, and the difference $\Delta = \sigma_i - \sigma_j$ is
always plotted such that $|b_i| \geq |b_j|$.  $\Delta_{\rm l,obs}$ and $\Delta_{\rm b,obs}$ denote the difference in observed proper motion dispersions  between adjacent fields in the longitude and latitude directions respectively. The equivalent quantities predicted from the model are denoted $\Delta_{\rm l,mod}$ and $\Delta_{\rm b,mod}$. In Fig.~\ref{fig:ObsDispersionResiduals} we see that the
deviations $\Delta_{\rm l,obs}$ and $\Delta_{\rm b,obs}$ scatter about 0, but have a quite broad distribution in both the $l$ and $b$ directions, with several fields inconsistent with
zero difference at $1 \sigma$ (defined as the sum in quadrature of the
uncertainties of the corresponding quantities of the two fields under
comparison).  Several deviations are as large as $0.2$ mas yr$^{-1}$
(corresponding to $\simeq 8 \kms$ at the Galactic centre) and many $\sigma$'s away from zero.  In view of the fact that these
differences have mean close to zero, it is possible that these
deviations are due to some systematic effect rather than to intrinsic
substructure in the Milky Way.  We return to this point briefly in the discussion.

\begin{figure}
\psfrag{ylabel1}{\normalsize \hspace{-25pt} $\Delta_{\rm l,mod}$ (\masyr)}
\psfrag{ylabel2}{\normalsize \hspace{-25pt} $\Delta_{\rm b,mod}$ (\masyr)}
\psfrag{xlabel}{\raisebox{-2pt}{\normalsize \hspace{-12pt} Separation $(^{\circ})$}}
\centering\includegraphics[width=1.0\hsize]{\FigDir{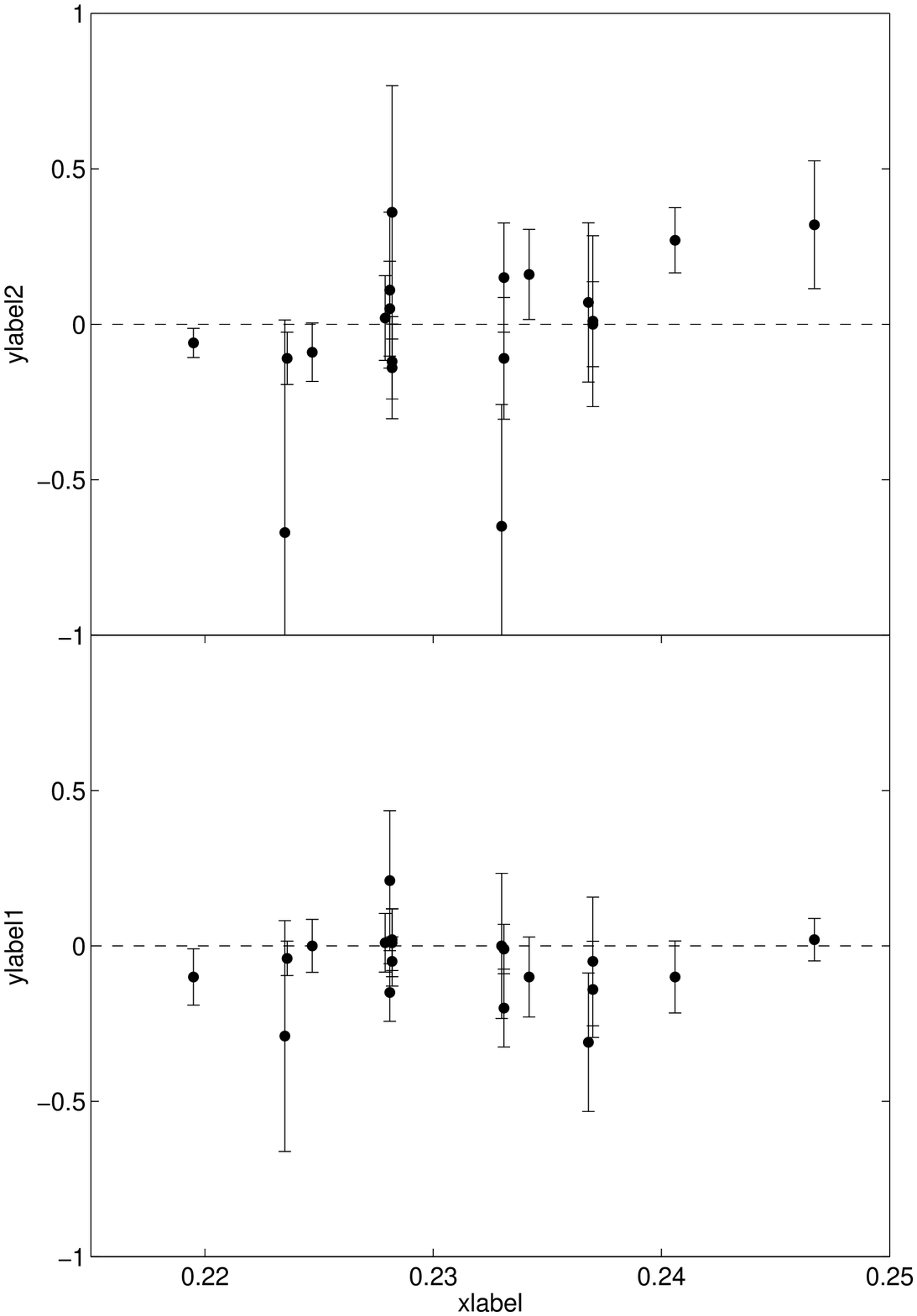}}
\caption{\label{fig:ModDispersionResiduals}Difference between model proper motion dispersions  for pairs of fields with separations less than 0.25 degrees.}
\end{figure}

In the case of the model uncertainties, however, Fig.~\ref{fig:ModDispersionResiduals}  shows that in most cases the differences $\Delta_{\rm l,mod}$ and $\Delta_{\rm b,mod}$
are consistent with zero at the $1 \sigma$ level, indicating that
these error estimates are robust.

\begin{figure}

\psfrag{xlabel}{\normalsize \raisebox{-2pt}{\hspace{-25pt} Galactic longitude $(\deg)$}}
\psfrag{ylabel1}{\normalsize \hspace{-10pt} $\delta_{\rm l}$ (\masyr)}
\psfrag{ylabel2}{\normalsize \hspace{-10pt} $\delta_{\rm b}$ (\masyr)}

\centering\includegraphics[width=1.0\hsize]{\FigDir{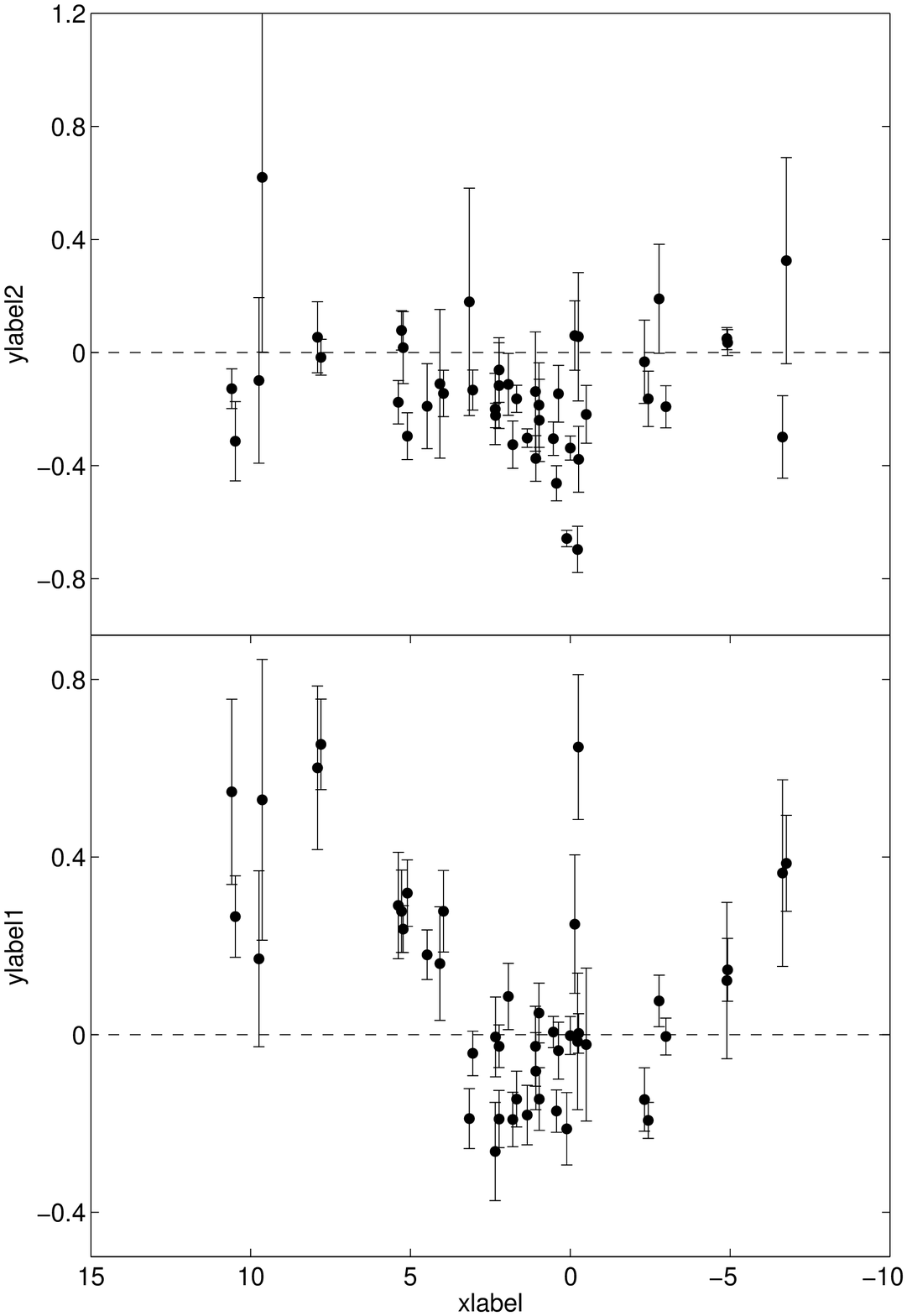}}
\caption{\label{fig:Residuals_l}Residuals $\delta_{\rm l,b} =
(\sigma_{\rm mod} - \sigma_{\rm obs})$ (see text), plotted against longitude,~$l$. }
\end{figure}

\begin{figure}
\psfrag{xlabel}{\normalsize \raisebox{-2pt}{\hspace{-25pt} Galactic latitude $(\deg)$}}
\psfrag{ylabel1}{\normalsize \hspace{-10pt} $\delta_{\rm l}$ (\masyr)}
\psfrag{ylabel2}{\normalsize \hspace{-10pt} $\delta_{\rm b}$ (\masyr)}

\centering\includegraphics[width=1.0\hsize]{\FigDir{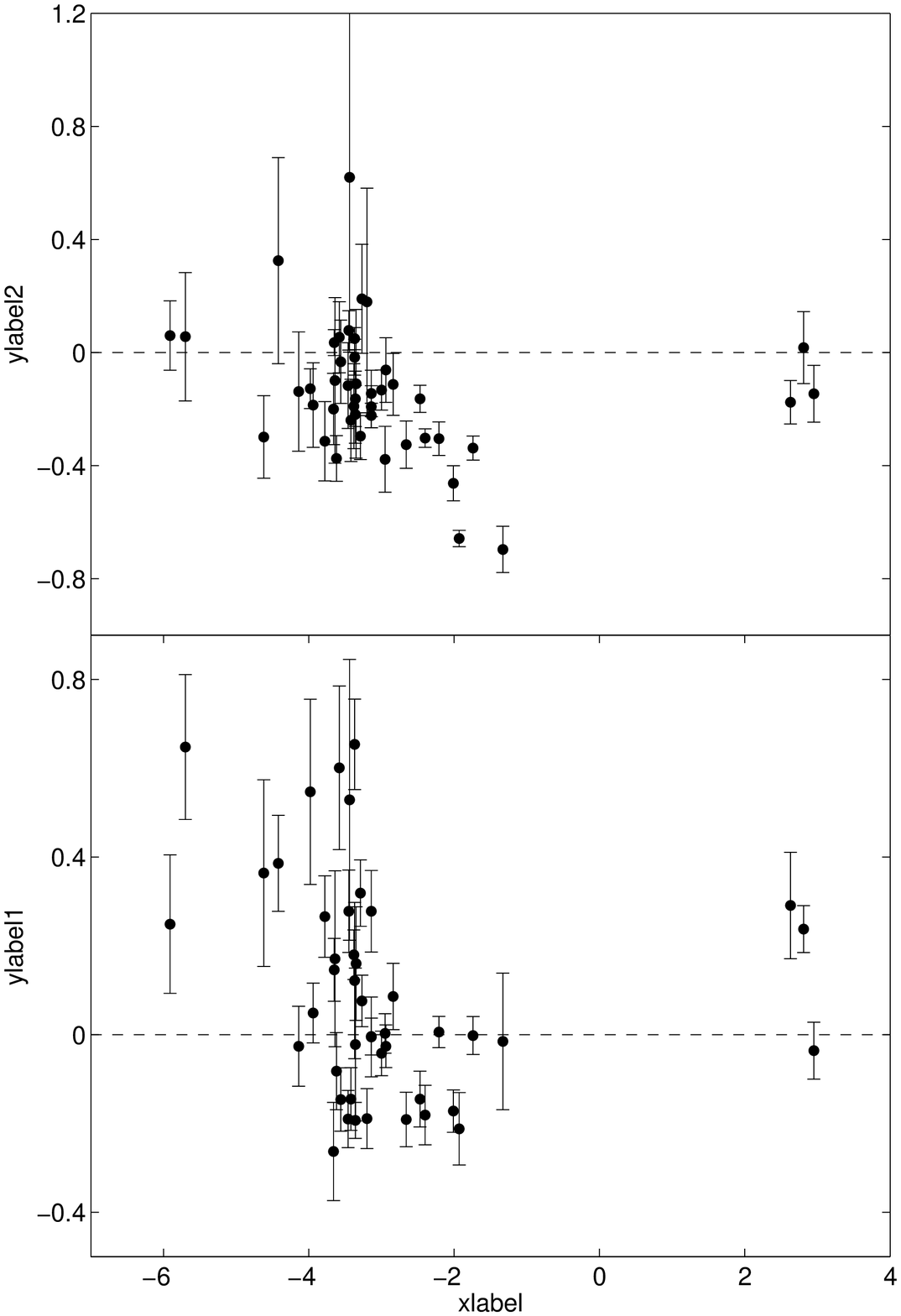}}
\caption{\label{fig:Residuals_b}Residuals $\delta_{\rm l,b} =
(\sigma_{\rm mod} - \sigma_{\rm obs})$ (see text), plotted against latitude,~$b$. }
\end{figure}

We now seek to explore the correlations of the residuals with
properties of the model.  We plot residuals $\delta_{\rm l,b} =
(\sigma_{\rm mod} - \sigma_{\rm obs})$, where $\sigma_{\rm mod}$ and $\sigma_{\rm obs}$ are the model and observed proper motion dispersions in the corresponding Galactic co-ordinate. The errorbar length is $(u_{\rm mod}^2 + u_{\rm obs}^2)^{1/2}$ where $u_{\rm mod}$ and $u_{\rm obs}$ are the uncertainties in the model and
observed proper motion dispersions, respectively.  Plotting these quantities as a
function of $l$, we note that there is no significant correlation, but
that the largest deviations in the latitude proper motion dispersion occur close to
$l=0$, see Fig.~\ref{fig:Residuals_l}.  In plotting $\delta_{\rm l,b}$ as a function of $b$, the reason which becomes evident is
that the fields closest to the mid-plane have the largest $\delta_{\rm b}$, see Fig.~\ref{fig:Residuals_b}. The density distribution in this region is uncertain due to presence of dust and the large extinction corrections required.  This may explain why the residuals of $\sigma_{\rm b}$ seem to correlate more with $b$ than those of $\sigma_{\rm l}$.  We note that the $\sigma_{\rm l}$ residuals also seem to have some dependence on $b$.  A possible explanation is that there is some additional effect due to dust which has not been accounted for.

\section{Discussion}
\label{sec:discussion}
Red clump giant stars in the dense fields observed by the OGLE-II
microlensing survey can be used as tracers of the bulge density and
motion over a large region toward the Galactic centre. The proper motion dispersions
of bulge RCG stars in the OGLE-II proper motion catalogue of
\citet{2004MNRAS.348.1439S} were calculated for 45 OGLE-II fields. The kinematics derived from the ground-based OGLE-II data were found to be in agreement with HST observations in two fields from \citet{2006MNRAS.370..435K}. It is reassuring that the results presented here are consistent with those derived from the higher resolution HST data, despite possible selection effects and blending.

The observed values of $\sigma_{\rm l}$ and $\sigma_{\rm b}$ were compared to
predictions from the made-to-measure stellar-dynamical model of
Debattista et al. (2007, in preparation).  In general, the model gives predictions qualitatively similar
to observed values of $\sigma_{\rm l}$ and $\sigma_{\rm b}$ for
fields close to the Galactic centre. The model is in agreement with
observed OGLE-II data in the direction previously tested by
\citet{2004ApJ...601L.155B}. Using the definition of \citet{DeL07}, the effective number of particles in the model used here is 3986. This relatively low number results in large errors on the model proper motion dispersions and we therefore recommend regarding interpretations based on this model with some caution.   An improved model with a larger number of particles (the recent study by \citet{DeL07} has an effective particle number $\sim 10^6$) will decrease the errors on the model predictions and allow a more useful comparison between model and observed proper motion dispersions. 

The OGLE-II fields mostly extend over
$\sim 17\deg$ in longitude and about $5\deg$ in latitude
across the Galactic bulge region and can therefore provide a more
powerful set of constraints on stellar motions predicted by galactic
models.  The high-accuracy proper motion data for the 45 fields and
those obtained with HST \citep{2006MNRAS.370..435K} can be used as
direct input in the made-to-measure method to construct a  better
constrained dynamical model of the Milky Way.

The statistical errors of our proper motion dispersions are small ($\sim
\kms$), but systematic uncertainties (for example due to incorrect
dust extinction treatment)  which were not included in the
analysis may be significant. Nevertheless, it is interesting to note
that there appears to be significant difference between the observed
proper motion dispersions of adjacent fields (e.g. fields 1 and 45). This might hint
at some fine-scale population effect,where the kinematics of
the bulge may be not in total equilibrium (e.g. due to a small
accretion event).  Higher-accuracy
observations using the HST may provide further evidence of such
population effects. We note that \cite{Rich06} suggest the possible existence of cold structures using data from a radial velocity survey of Galactic bulge M giant stars although their conclusion could be strengthened by a larger sample of stars.

The OGLE-II proper motion catalogue \citep{2004MNRAS.348.1439S} for millions of bulge stars is still somewhat under-explored. For example, it will be interesting to explore the nature of the high proper motion stars ($\mu >10$ \masyr) and search for wide binaries in the catalogue. Some exploration along these lines is under way.

\section*{Acknowledgements}

We thank Drs. Vasily Belokurov, Wyn Evans and Martin Smith for helpful
discussions, and the anonymous referee for their helpful suggestions.

NJR acknowledges financial support by a PPARC PDRA fellowship.  This
work was partially supported by the European Community's Sixth
Framework Marie Curie Research Training Network Programme, Contract
No. MRTN-CT-2004-505183 `ANGLES'.   VPD is supported by a Brooks
Prize Fellowship at the University of Washington and receives partial 
support from NSF ITR grant PHY-0205413.


\end{document}